\documentclass{article}

\usepackage{arxiv}

\usepackage[utf8]{inputenc} % allow utf-8 input
\usepackage[T1]{fontenc}    % use 8-bit T1 fonts
\usepackage{hyperref}       % hyperlinks
\usepackage{url}            % simple URL typesetting
\usepackage{booktabs}       % professional-quality tables
\usepackage{amsfonts}       % blackboard math symbols
\usepackage{nicefrac}       % compact symbols for 1/2, etc.
\usepackage{microtype}      % microtypography
\usepackage{lipsum}
\usepackage{color}
\usepackage{algorithm}
\usepackage{algorithmic}
\usepackage{amsmath,amsfonts,amssymb}
\usepackage{subfigure}
\usepackage{lineno}
\usepackage{graphicx}
\usepackage{bm}
\graphicspath{{Figures/}}
\usepackage{subfigure}

\title{Data-driven transient lift attenuation\\
for extreme vortex gust-airfoil interactions}

\author{
Kai Fukami$^{[1,*]}$, Hiroya Nakao$^{[2]}$, Kunihiko Taira$^{[1]}$
\\
\\
1. Department of Mechanical and Aerospace Engineering,
University of California, Los Angeles, CA 90095, USA\\
2. Department of Systems and Control Engineering, Tokyo Institute of Technology, Tokyo, 152-8550, Japan\\
Corresponding author: kfukami1@g.ucla.edu
}

\begin{document}
\maketitle

\begin{abstract}

We present a data-driven feedforward control to attenuate large transient lift experienced by an airfoil disturbed by an extreme level of discrete vortex gust.
The current analysis uses a nonlinear machine-learning technique to compress the high-dimensional flow dynamics onto a low-dimensional manifold.
While the interaction dynamics between the airfoil and extreme vortex gust are parameterized by its size, gust ratio, and position, the wake responses are well-captured on this simple manifold.
The effect of extreme vortex disturbance about the undisturbed baseline flows can be extracted in a physically-interpretable manner.
Furthermore, we call on phase-amplitude reduction to model and control the complex nonlinear extreme aerodynamic flows.
The present phase-amplitude reduction model reveals the sensitivity of the dynamical system in terms of the phase shift and amplitude change induced by external forcing with respect to the baseline periodic orbit.
By performing the phase-amplitude analysis for a latent dynamical model identified by sparse regression, the sensitivity functions of low-dimensionalized aerodynamic flows for both phase and amplitude are derived.
With the phase and amplitude sensitivity functions, optimal forcing can be determined to quickly suppress the effect of extreme vortex gusts towards the undisturbed states in a low-order space.
The present optimal flow modification built upon the machine-learned low-dimensional subspace quickly alleviates the impact of transient vortex gusts for a variety of extreme aerodynamic scenarios, providing a potential foundation for flight of small-scale air vehicles in adverse atmospheric conditions.

\end{abstract}

\section{Introduction}
\label{sec:intro}

Small-scale air vehicles are used in a range of operations including transportation~\cite{cai2014survey}, rescue~\cite{mishra2020drone}, agriculture~\cite{zhang2012application} and broadcasting~\cite{holton2015unmanned}.
Although such small-scale aircraft typically fly in calm air, they are now being tasked to navigate in challenging environments such as urban canyons, mountainous areas, and turbulent wakes created by ships.
As the occurrence of these extreme scenarios has increased due to climate change, reliable control strategies are critical to achieving stable flight in violent atmospheric disturbances~\cite{jones2022physics,mohamed2023gusts}.
In response, this study presents a data-driven flow control approach for a wing experiencing extreme levels of vortical gusts.

In violently adverse airspace, small-scale air vehicles encounter various forms of vortex disturbance characterized by a number of parameters including its vortex strength, size, and orientation~\cite{biler2021experimental,stutz2023dimensional}.
In studying vortex-gust airfoil interaction, the gust ratio $G  \equiv u_g/u_\infty$ is a particularly important factor, where $u_g$ is the characteristic gust velocity and $u_\infty$ is the freestream velocity or cruise velocity.
Flight condition of $G > 1$ is traditionally avoided, which can occur in urban canyons, mountainous environments, and severe atmospheric turbulence~\cite{jones2022physics,jones2021overview}.
Large-scale aircraft generally do not encounter conditions of $G > 1$ due to their high-cruise velocity.
However, such a condition becomes an important concern for small-scale aircraft such as drones because of its low cruise velocity, leading to potentially large $G$.

Considering such severe conditions in which the spatiotemporal scales of the baseflow unsteadiness and disturbances reach almost the same level in magnitude, our recent study has examined extremely high levels of aerodynamic disturbances with $0<G\leq 10$~\cite{FT2023}.
In particular, we refer to aerodynamics with $G>1$ as {\it extreme aerodynamics} due to the presence of violently strong gusts.

Previous studies of vortex-gust airfoil interactions have mainly focused on scenarios with $G \leq 1$.
For example, Qian et al.~\cite{qian2023lift} experimentally investigated vortex-gust airfoil interaction under $G \leq 0.5$.
They examined the effect of various parameters such as gust ratio, angle of attack, and sweep angle of the wing on vortical flows and aerodynamic forces through PIV measurements.
Herrmann et al.~\cite{herrmann2022gust} considered gust mitigation of flows around a DLR-F15 airfoil under vortex gusts with $G \leq 0.1$.
With trailing-edge flaps and a combined proportional-integral feedback/model-based feedforward approach, they achieved 64\% reduction in the lift deviation during quasirandom gust encounters.
For conditions of $G \leq 0.71$, Sedky et al.~\cite{sedky2023experimental} has recently developed a closed-loop pitch control strategy to mitigate lift fluctuation for transverse gust encounters.

Our recently-proposed data-driven technique called a nonlinear lift-augmented autoencoder uncovers the low-dimensional dynamics of vortical flows experiencing extreme levels of vortex disturbances over a wide parameter space~\cite{FT2023}.
We have found that time-varying vortical flow fields spanning the large parameter space can be compressed to only three variables using nonlinear machine learning.
In the latent space composed of the three variables, the dynamical trajectories converge to a certain structure, forming the low-dimensional inertial manifold that captures the influence of extreme vortex disturbance on the baseline flow dynamics.

\begin{figure}
    \centering
    \includegraphics[width=1\textwidth]{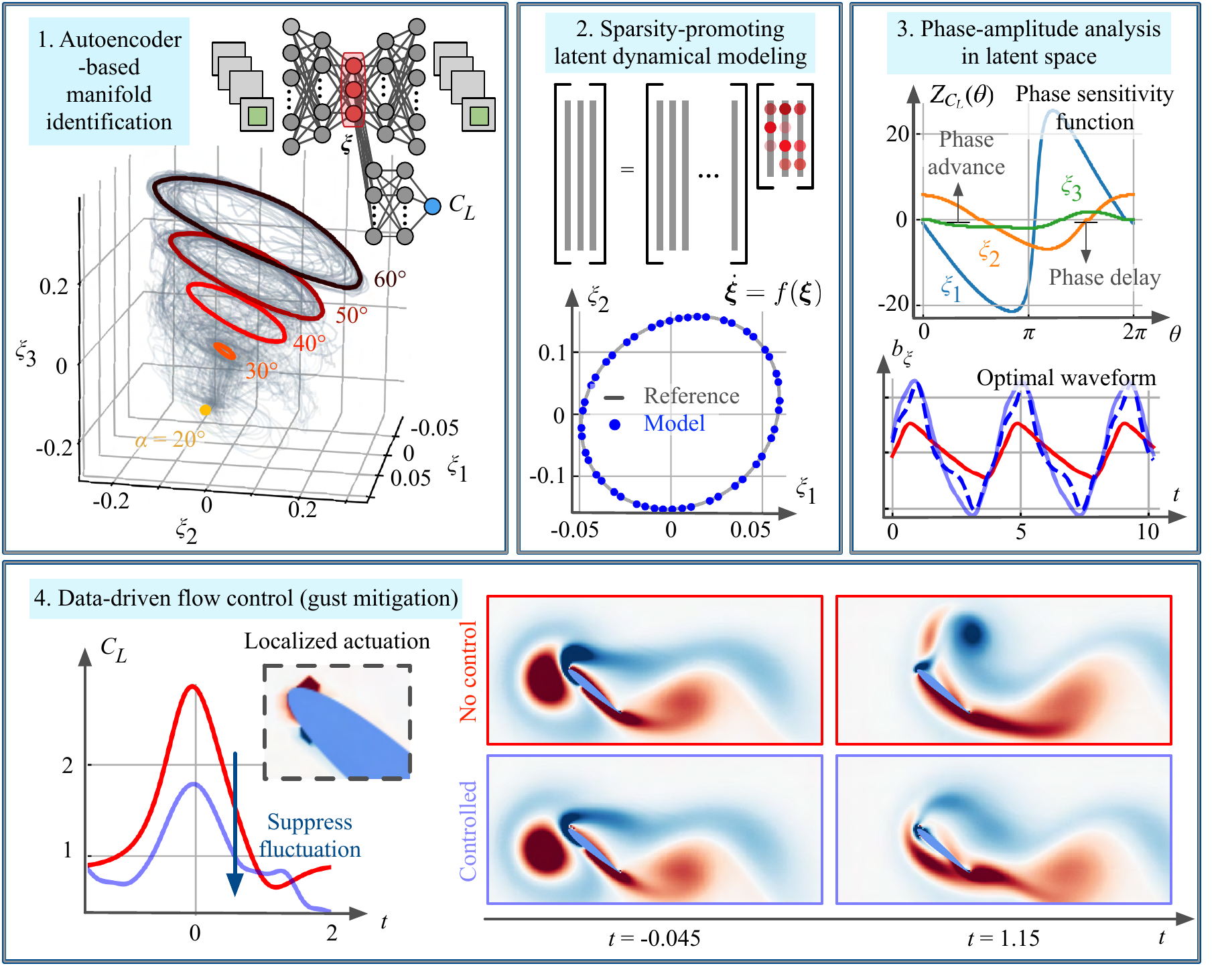}
    \vspace{-5mm}
    \caption{
    Overview of the present study: nonlinear data compression (section~\ref{sec:FT2023}), dynamical modeling (sections~\ref{sec:SINDy} and~\ref{sec:SINDyresult}), control design with phase-amplitude reduction (sections~\ref{sec:PRA} and~\ref{sec:PAresult}), and flow control (sections~\ref{sec:AFE} and~\ref{sec:cont_result}).
    }
    \vspace{-3mm}
    \label{fig1}
\end{figure}

The complex dynamics of vortex-airfoil interactions are driven not only by the gust ratio but also by other factors such as the Reynolds number, wing geometry, disturbance size, and orientation.
Since different combinations of these parameters create diverse patterns of vortex-airfoil interactions, covering infinitely different scenarios with numerical and experimental studies by na\"ive parameter sweeps is impractical.
This calls for a smart way to sample and extract the fundamental nonlinear dynamics.
There is also a need to control these violent flows to achieve some form of stable flight.

For achieving real-time control, a reduced-order model is necessary.
Linear techniques such as proper orthogonal decomposition~\cite{lumley1967structure,berkooz1993proper} and dynamic mode decomposition~\cite{Schmid2010} have been used to extract low-dimensional flow features and dynamics.
However, finding a universal low-order representation over a range of flow configurations or patterns is challenging with linear techniques when mode deformation occurs.
In such a case, nonlinear machine-learning-based techniques can be helpful~\cite{BEF2019,BHT2020,BNK2020}.

This study considers leveraging the machine-learned low-order manifold for gust mitigation control.
However, controlling such violent flows is challenging due to their transient nature.
To address this point, we apply phase-amplitude reduction~\cite{nakao2021phase,shirasaka2017phase} to the low-dimensional extreme aerodynamic manifold for the design of a control law.
Phase-amplitude reduction is a technique to analyze oscillatory signals or waveforms in a range of nonlinear dynamic problems~\cite{wedgwood2013phase,wilson2016isostable,YFTN2024}.
This analysis can model a given complex dynamics with its phase and amplitude.
Phase can be thought of as the timing information of a signal, referring to the position of a waveform at a particular point over time relative to a reference point.
On the other hand, amplitude represents the intensity of the deviation of the waveform from the reference at a specific point in time~\cite{mauroy2018global,kotani2020nonlinear,MZN2023}.

A simplified form of a given complex dynamics with the reduction to its phase and amplitude facilitates dynamical modeling and system control~\cite{nakao2016phase,kurebayashi2013phase,mauroy2013isostables,takeda2023two}.
Phase-reduction analysis has recently been used to characterize and control fluid flows, including the periodic vortex shedding around cylinders~\cite{taira2018phase,iima2019jacobian,khodkar2020phase,khodkar2021phase,loe2021phase,loe2023controlling}, a flat plate~\cite{iima2021phase,iima2023optimal}, and airfoil~\cite{nair2021phase,asztalos2021modeling,kawamura2022adjoint,godavarthi2023optimal}.
Synchronization characteristics to various forms of periodic perturbations in fluid flows can also be examined with phase-reduction analysis, demonstrated with vortex shedding for a circular cylinder~\cite{taira2018phase,khodkar2020phase,khodkar2021phase,nair2021phase}.
For laminar-separated airfoil wakes, phase-reduction-based control design has also shown promise not only to reveal responsible flow physics~\cite{kawamura2022adjoint} but also to optimally modify the wake dynamics~\cite{godavarthi2023optimal}.

This study develops a feedforward control strategy to quickly mitigate the impact of an extreme discrete vortex gust by leveraging the phase-amplitude reduction model on the extreme aerodynamic manifold.
The overview of this study is presented in figure~\ref{fig1}.
There is a step-by-step procedure for preparing the optimal control actuation, aiming to quickly modify the flow state.

The present paper is organized as follows.
Extreme aerodynamic flow physics and their low-dimensionalization through a machine-learning technique are introduced in section~\ref{sec:FT2023}.
The method used to prepare the optimal control strategy is described in section~\ref{sec:method}.
Results are presented in section~\ref{sec:res_dis}.
At last, conclusions are offered in section~\ref{sec:conc}.

\section{Extreme vortex-airfoil interactions on a low-dimensional manifold}
\label{sec:FT2023}

\begin{figure}[b]
    \centering
    \includegraphics[width=0.9\textwidth]{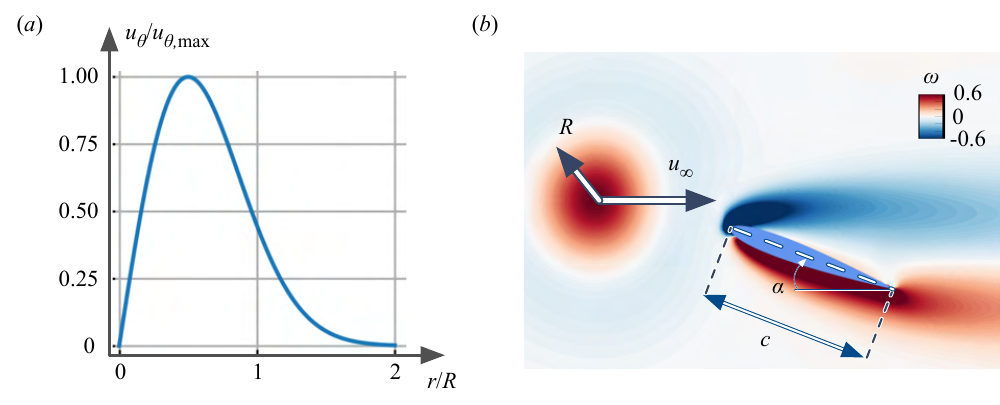}
    \vspace{-2mm}
    \caption{
    $(a)$ The velocity profile of the vortex gust.
    $(b)$ An example vorticity field with a vortex gust. 
    The parameters considered in the present study are also shown.
    The same color scale of vorticity field visualization is hereafter used throughout the paper.
    }
    \vspace{-3mm}
    \label{fig2}
\end{figure}

In this study, we consider extreme vortex-gust airfoil interactions that exhibit strong transient and nonlinear dynamics.
To control such violent aerodynamic flows, we develop a data-driven strategy using {the sparse identification of nonlinear dynamics (SINDy)~\cite{brunton2016discovering} and phase-amplitude reduction analysis~\cite{takata2021fast} on a low-dimensional manifold, as presented in figure~\ref{fig1}.
This section first introduces the model problem and discusses the complex transient flow physics of extreme aerodynamics.
We then show how complex, high-dimensional vortical flows under extreme aerodynamic conditions can be compactly expressed in the latent space using nonlinear autoencoder.
{Sections~\ref{subsec:AE} and~\ref{subsec:manifold} present the current autoencoder formulation and its use for identifying low-dimensional representations of the dynamics~\cite{FT2023}.}

% The first step is to obtain a universal, low-dimensional representation of extreme aerodynamic flows.
% This is achieved by a nonlinear machine-learning-based technique referred to as a lift-augmented autoencoder~\cite{FT2023}.
% Once we obtain the low-dimensional expression, we model the latent dynamics using sparsity-promoting regression.
% The identified dynamical model here is needed to perform phase-amplitude reduction.
% Through the phase-amplitude reduction, we can measure phase- and amplitude sensitivity functions that are used to derive a control law.
% In this study, we aim to achieve significant control effects within a very short time duration because the impact of the gust quickly generally appears within only 2--3 convective time.
% The phase-amplitude reduction enables us to optimally gain a control actuation that can quickly modify the flow state while suppressing the amplitude modulation of the dynamics.
% The derived time-varying control actuation is finally applied to extreme aerodynamic flows.

\begin{figure}[H]
    \centering
    \includegraphics[width=\textwidth]{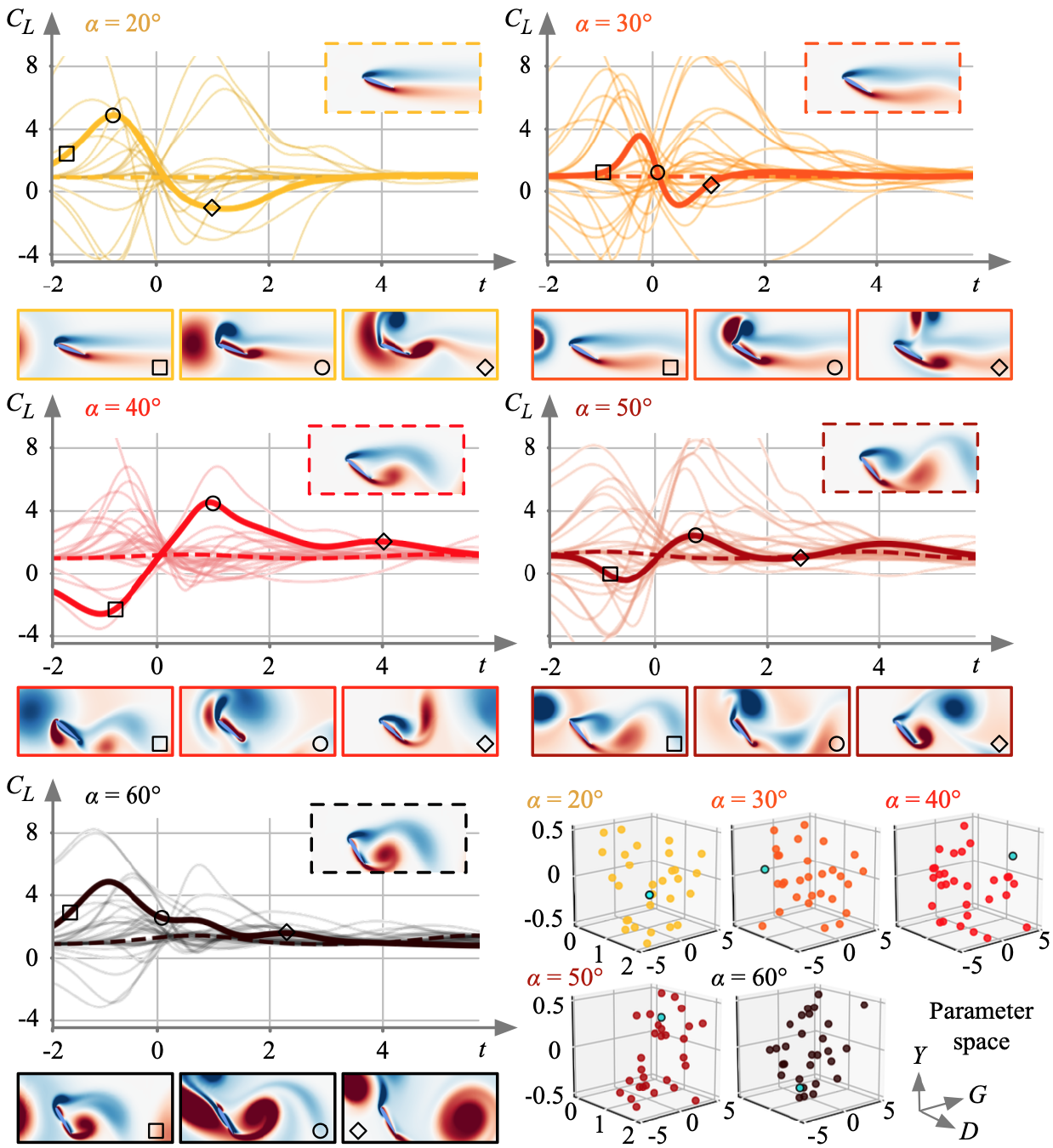}
    \vspace{-2mm}
    \caption{
    % Extreme aerodynamic flow data set.
    The entire collection of lift history over the parameter space of $(\alpha,G,D,Y)$ with representative vorticity fields.
    The vorticity field surrounded by the box (dashed line) is the undisturbed flow for each angle of attack.
    % A time series of representative disturbed vorticity fields surrounded by the solid line is also visualized.
    The dashed and solid lines in the lift curve correspond to the undisturbed case and a representative disturbed case, respectively.
    The light-blue circles in the parameter spaces correspond to the representative cases chosen for the vorticity field visualizations.
    }
    \vspace{-3mm}
    \label{fig3}
\end{figure}

\subsection{Flow physics: extreme vortex-gust airfoil interaction}

As a model problem, we consider extreme vortex-gust airfoil interaction around a NACA0012 airfoil with angles of attack $\alpha \in [20,60]^\circ$ at a chord-based Reynolds number of 100.
% in which the wing experiences a significant level of transient force.
The data sets are produced by {fully-resolved} (direct) numerical simulations~\cite{cliff1,cliff2,FT2023}.
The computational domain is set over $(x, y)/c \in [-15, 30] \times [-20, 20]$ with the leading edge of the wing positioned at the origin. 
Verification and validation have been performed extensively with previous studies~\cite{ZFAT2023,kurtulus2015unsteady,liu2012numerical,di2018fluid}.

Without the presence of a vortex gust, a wake at $\alpha = 20^\circ$ is steady while wakes at $\alpha \geq 30^\circ$ exhibit unsteady periodic shedding (limit-cycle oscillation).
{The current high $\alpha$ is motivated to model unsteady operating (base) conditions at the present Reynolds number.}
For the disturbed wake cases, an extremely strong vortex gust is introduced upstream of a wing at $x_0/c = -2$ and $y_0/c\equiv Y \in [-0.5, 0.5]$, as illustrated in figure~\ref{fig2}.
A Taylor vortex~\cite{taylor1918dissipation} is used to model the disturbance with a rotational velocity profile of
\begin{align}
    u_{\theta}=u_{\theta, \rm{max}}{\dfrac{r}{R}}{\rm exp}\left[\dfrac{1}{2}\left({1-\dfrac{r^2}{R^2}}\right)\right] ,
\end{align}
where $R$ is the radius at which $u_{\theta}$ reaches its maximum velocity $u_{\theta, \rm{max}}$.
{To cover a variety of wake patterns in this study, the vortex gust is created by randomly chosen} gust ratio $G \equiv u_{\theta,\text{max}}/u_\infty \in [-4,4]$, its size $D \equiv 2R/c \in [0.5, 2]$, and vertical position of the disturbance $Y$ relative to the wing.
Note that the range of gust ratio $G$ considered herein is much larger than that traditionally thought of as flyable~\cite{jones2022physics}.

\begin{figure}[H]
    \centering
    \includegraphics[width=\textwidth]{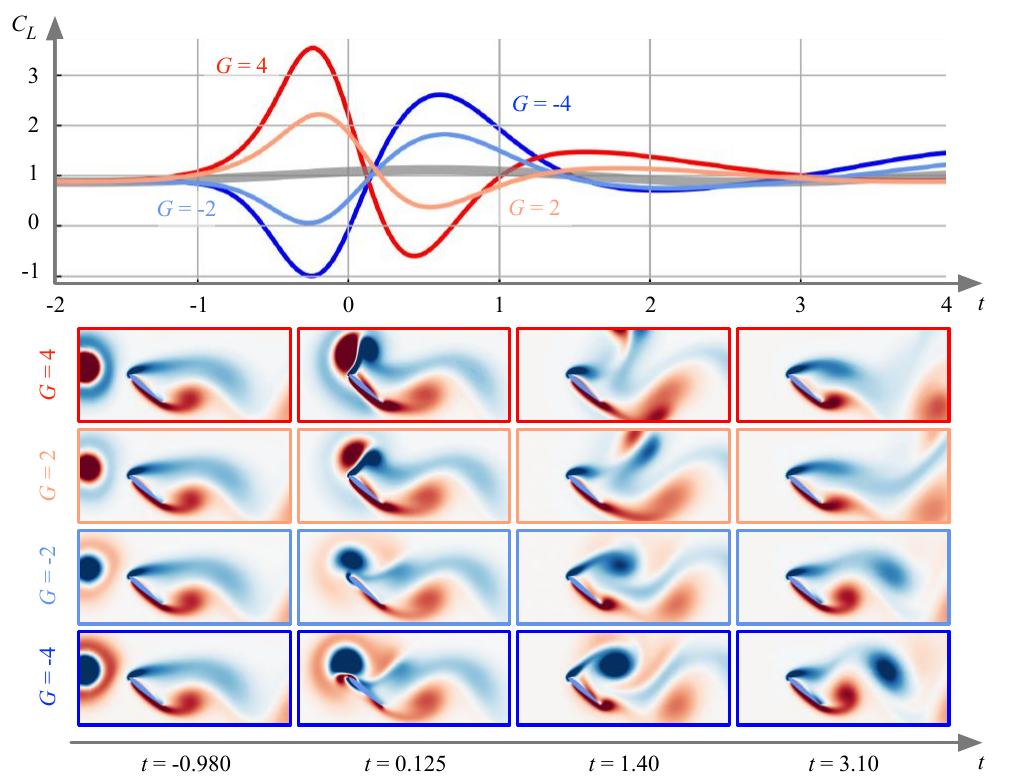}
    \vspace{-4mm}
    \caption{
    Dependence of lift coefficient $C_L$ and vorticity field $\bm \omega$ on the gust ratio $G$.
    The cases for $(\alpha,D,Y)=(40^\circ,0.5,0.1)$ with $G=\pm 2$ and $\pm 4$ are shown.
    The gray line in the lift response corresponds to the baseline (undisturbed) case.
    }
    \vspace{-3mm}
    \label{fig4-new}
\end{figure}

% A parameter combination of the vortex-gust setting generates diverse patterns of violent vortex-airfoil interaction~\cite{FT2023}.
Let us exhibit in figure~\ref{fig3} the entire collection of lift responses in the present data with representative vortical field snapshots.
Here, the convective time is set to zero when the center of the vortex arrives at the leading edge of the airfoil.
{The present dataset includes 150 disturbed flow cases with 30 cases for each angle of attack.}
% The present data set includes a variety of wake behaviors and lift waveforms over time, occurring due to complex vortex-gust airfoil interactions.
Strong vortex gusts induce a large excitation of aerodynamic forces within a very short time with highly nonlinear transient flow dynamics.
Furthermore, the flow exhibits a variety of wake patterns depending on the parameter combination of the setup including the vortex size, strength, and initial position.
Due to the nonlinear interaction between the vortex gust and a flow around an airfoil, massive flow separation can occur, creating additional vortical structures.
% This study aims to mitigate the impact of such a strong vortex gust with data-driven control.

\begin{figure}
    \centering
    \includegraphics[width=\textwidth]{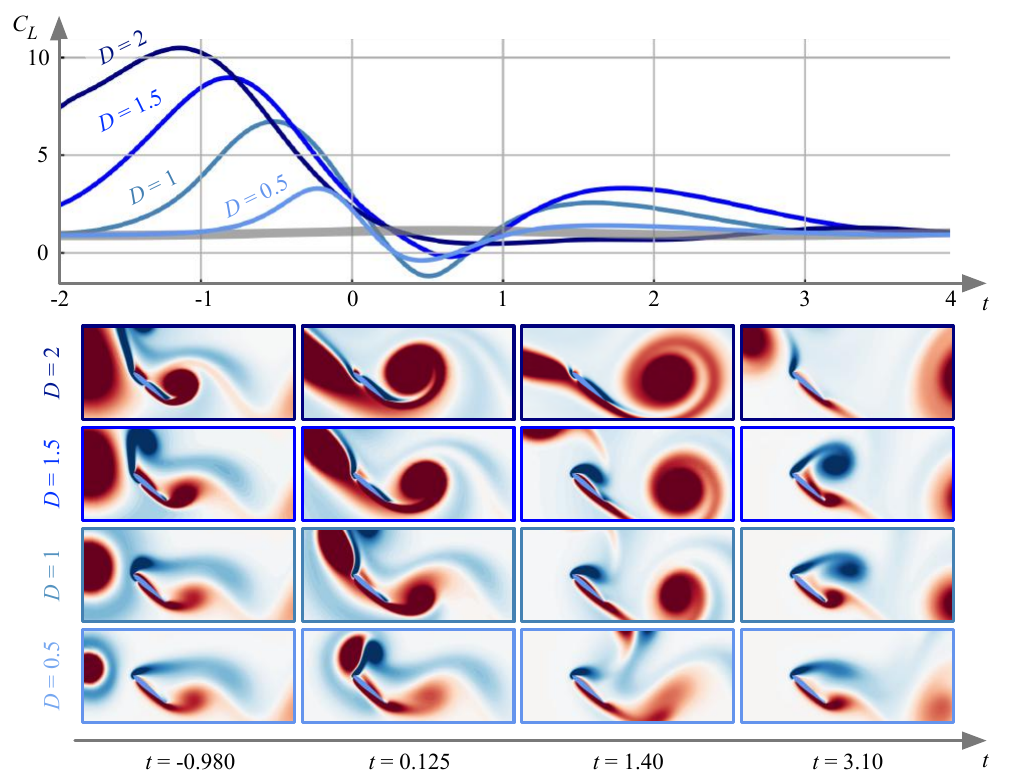}
    \vspace{-4mm}
    \caption{
    Dependence of lift coefficient $C_L$ and vorticity field $\bm \omega$ on the gust size $D$.
    The cases for $(\alpha,G,Y)=(40^\circ,3.6,0.1)$ with $D=0.5$, 1, 1.5, and 2 are shown.
    The gray line in the lift response corresponds to the baseline (undisturbed) case.
    }
    \vspace{-3mm}
    \label{fig6-new1}
\end{figure}

The gust ratio $G$ is one of the critical parameters that influence the flyability of air vehicles. 
Here, we examine the effect of $G$ on lift coefficient $C_L$ and vorticity field $\bm \omega$ for cases of $(\alpha,D,Y)=(40^\circ,0.5,0.1)$, as shown in figure~\ref{fig4-new}.
We consider $G=\pm 2$ and $\pm 4$ as representative examples.
For the positive vortex gusts, lift first increases from the undisturbed state.
Once the positive vortex gust impinges on the airfoil, the interaction between the gust and the airfoil wake causes massive separation, contributing to the decrease of the lift over $0<t<1$. 
In contrast, negative vortex disturbances decrease the lift first with subsequent lift value recovery towards that of the original limit-cycle case in a transient manner.
Note that the transient lift generated by these vortices is very large compared to the undisturbed lift level.
The fluctuation from the undisturbed lift generally increases as $|G|$ becomes large.

{It is also observed that the difference in $G$ of the positive gust cases causes the shift in timing for the secondary peak of $C_L$ from $t\approx 0.5$ ($G=2$) to $0.6$ ($G=4$).}
This is because the gust with $G=4$ interacts with the pre-existing negative vorticity on the suction side of the airfoil more strongly than that with $G=2$.
As depicted with the vorticity snapshots at $t=0.125$, a stronger interaction with $G=4$ forms a larger negative vortex near the leading edge, compared to the case with $G=2$.

The dependence of the extreme aerodynamic response on the gust size is also investigated, as shown in figure~\ref{fig6-new1}. 
For comparison, we fix the angle of attack, gust ratio, and vertical position $(\alpha, G, Y)=(40^\circ, 3.6, 0.1)$ while varying the gust size $D$ from 0.5 to 2.
For all the disturbed cases with different $D$, the lift response exhibits the same trend of first increasing and then decreasing towards the original undisturbed case.
The first lift peak appears earlier as the gust size increases since a larger vortex gust reaches the wing earlier, as presented in figure~\ref{fig6-new1}.
While the gust of $D=0.5$ primarily interacts with the structures near the leading edge, the vortex gust of $D>1$ simultaneously impacts the leading and trailing edge vortices, exhibiting massive separation while newly generating large vortical structures.

The extreme vortex-airfoil interaction dynamics are also affected by the vortex position $Y$ in addition to gust ratio $G$ and gust size $D$. 
This causes the difference in the interaction of a vortex gust with the pre-existing vortical structures around a wing.
To examine this point, we cover three vertical positions of $Y = (-0.3, 0, 0.3)$ for fixed parameters of $(\alpha, G, D)=(40^\circ, -2.2, 0.5)$, as presented in figure~\ref{fig5-new1}.
The lift fluctuation for $Y=0.3$ from the undisturbed case is smaller than that for $Y=0$ and -0.3 since only the bottom half of the negative vortex gust impinges the airfoil.
By shifting the vortex position downward, a large portion of the gust interacts with the airfoil, producing a large variation of lift force.
For $Y=-0.3$, the wing is largely affected by the negative vortex gust at the pressure side, experiencing a larger drop in lift force compared to the other two scenarios.

\begin{figure}
    \centering
    \includegraphics[width=\textwidth]{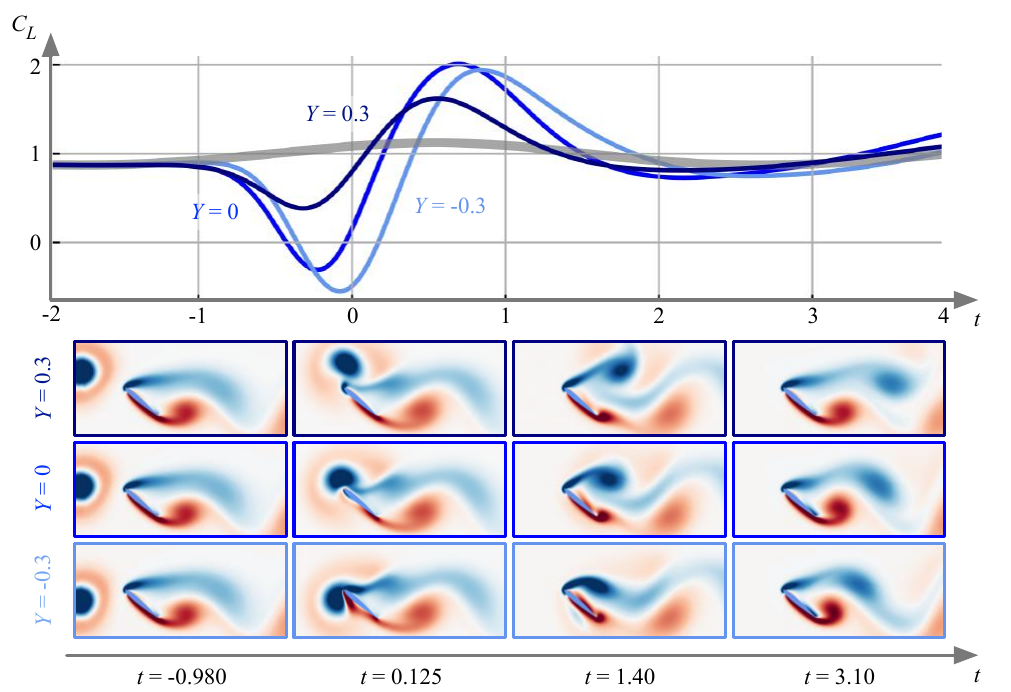}
    \vspace{-4mm}
    \caption{
    Dependence of lift coefficient $C_L$ and vorticity field $\bm \omega$ on the initial vertical position $Y$.
    The cases for $(\alpha,G,L)=(40^\circ,-2.2,0.5)$ with $Y=-0.3$, 0, and 0.3 are shown.
    The gray line in the lift response corresponds to the baseline (undisturbed) case.
    }
    \vspace{-3mm}
    \label{fig5-new1}
\end{figure}

We further note that the sharp lift responses from extreme vortex-gust airfoil interaction discussed above occur only within two convective times for almost all considered cases.
While we easily recognize the difficulty of controlling air vehicles under such a significant variation in the lift force, it also implies that a controller for the present extreme aerodynamic flows needs to quickly modify the flow to attenuate the transient lift responses.
This calls for a control technique that can react fast.

\subsection{Lift-augmented nonlinear autoencoder}
\label{subsec:AE}

Analyzing the present extreme aerodynamic flows is challenging due to their complexity and nonlinearity.
Furthermore, it is {challenging} to perform a large number of numerical simulations or experiments for studying vortex-airfoil interaction across a large parameter space with finite resources.
Hence, a model that universally captures the fundamental physics of extreme aerodynamics without necessitating expensive simulations and experiments would be beneficial.

In response, we have recently developed a lift-augmented nonlinear autoencoder~\cite{FT2023} that can compress a collection of extreme aerodynamic vortical flow data across a large parameter space into only few latent space variables while retaining the original vortex-airfoil interaction.
An autoencoder is a neural-network-based model reduction technique~\cite{HS2006}.
As illustrated in figure~\ref{fig4}, an autoencoder is composed of an encoder ${\cal F}_e$ and a decoder ${\cal F}_d$ while having the bottleneck where the latent vector ${\bm \xi}$ is positioned.
The autoencoder model is generally trained to output the same data as a given input data.
In other words, a given high-dimensional input data can be compressed into the latent vector $\bm \xi$ {if the autoencoder can successfully decode the original data.}

\begin{figure}
    \centering
    \includegraphics[width=0.75\textwidth]{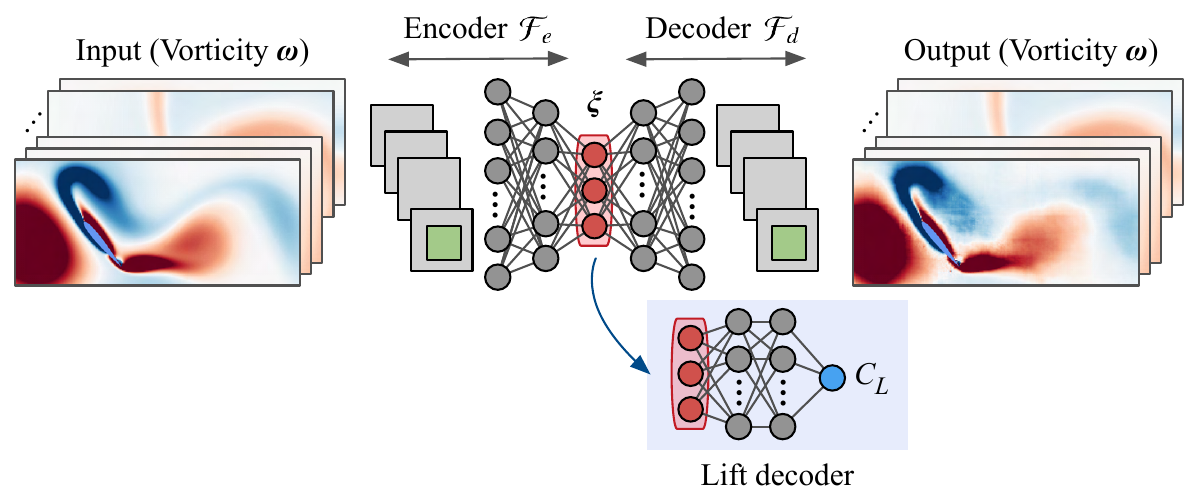}
    % \vspace{-2mm}
    \caption{
    Lift-augmented nonlinear autoencoder~\cite{FT2023}.
    }
    % \vspace{-3mm}
    \label{fig4}
\end{figure}

% ${\cal F}(\bm \omega) = {\cal F}_d({\cal F}_e({\bm \omega})) \approx {\bm \omega}$

In this study, the discrete vorticity field $\bm \omega$ is compressed through the autoencoder such that 
\begin{align}
    {\bm \omega}\approx{\cal F}(\bm \omega) = {\cal F}_d({\cal F}_e({\bm \omega})),
    ~~~
    {\bm \xi} = {\cal F}_e(\bm \omega),~~~ \bm\omega \approx \hat{\bm\omega} = {\cal F}_d({\bm \xi}),
\end{align}
where $\hat{\bm\omega}$ is the decoded (reconstructed) vorticity field.
The weights ${\bm w}$ inside a regular autoencoder are optimized by solving the following minimization problem, 
\begin{equation}
    {\bm w}^* = {\rm argmin}_{\bm w} \| {\bm\omega} - \hat{\bm\omega} \|_2={\rm argmin}_{\bm w} \| {\bm\omega} - {\cal F}({\bm\omega; {\bm w}}) \|_2,~\label{eq:ae}
\end{equation}
where ${\bm w}$ is the weights of the autoencoder.
By using nonlinear activation functions inside the neural network, an autoencoder can nonlinearly compress high-dimensional data into a low-order subspace, which often achieves higher compression than linear techniques.

While nonlinear autoencoders can be used to compress a variety of vortical flow data~\cite{fukami2021model,omata2019,xu2020multi,racca2023predicting,smith2024cyclic}, we have found that the regular formulation expressed in equation~\ref{eq:ae} does not produce a physically-interpretable data distribution in the latent space~\cite{FT2023}.
Extracting low-order coordinates associated with dominant aerodynamic features is important in considering not only {the interpretation of} extreme aerodynamic flows but also downstream tasks such as developing control strategies.
To facilitate the identification of a low-dimensional subspace from the aspect of aerodynamics, the proposed model referred to as a lift-augmented nonlinear autoencoder incorporates the lift coefficient $C_L(t)$.

In the present formulation, the additional branch network connected with the latent variables ${\bm \xi}(t)$ (lift decoder, the blue-shade portion in figure~\ref{fig4}) outputs $C_L(t)$.
This side network enables ${\bm w}$ to be tuned to capture important vortical structures that are correlated lift due to $\Gamma \propto C_L$, where $\Gamma$ is circulation.
This augmentation also helps to capture large transient lift caused by the present extreme vortex-airfoil interactions. 
The optimization for the weights inside the lift-augmented autoencoder is performed with
\begin{align}
    {\bm w}^* = {\rm argmin}_{\bm w}\biggl[||{\bm \omega}-\hat{\bm \omega}||_2 + \beta ||C_L - \hat{C_L}||_2 \biggr],
\end{align}
where $\beta$ balances the vorticity field and lift reconstruction loss terms.
{Here, the weights inside the main part and lift decoder are simultaneously optimized.}
For the present data-driven study, we use 1200 vorticity snapshots over 10.2 convective time per case.
A subdomain of $(x, y)/c \in [-1.4, 4] \times [-1.2, 1.2]$ with spatial grid points $(N_x, N_y) = (240, 120)$ is extracted from the computational domain for the machine-learning analysis. 
Details on this autoencoder formulation are provided in Fukami \& Taira~\cite{FT2023}.

\subsection{Vortex-airfoil interaction on a low-dimensional manifold}
\label{subsec:manifold}

\begin{figure}
    \centering
    \includegraphics[width=0.85\textwidth]{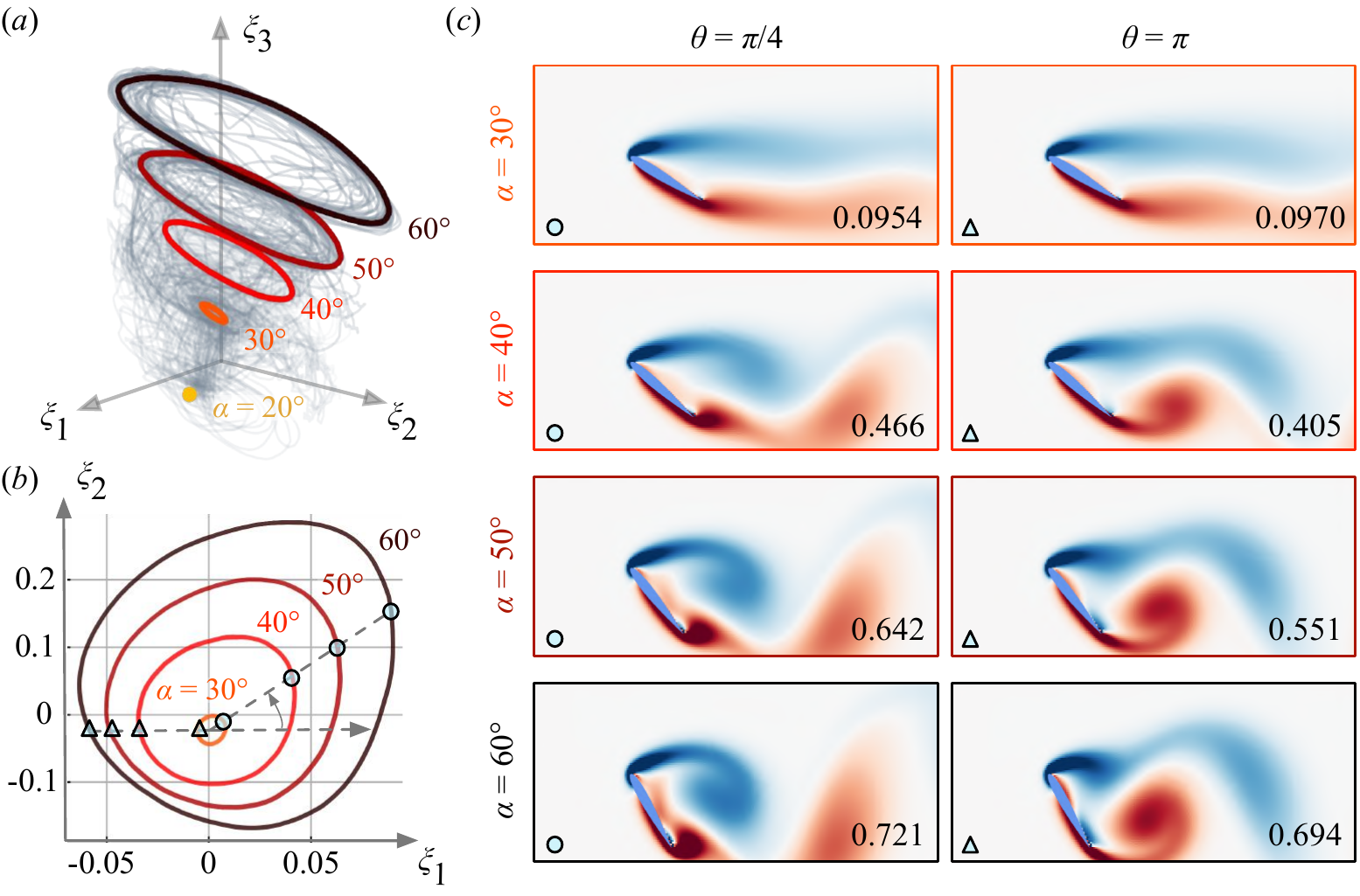}
    % \vspace{-2mm}
    \caption{
    Extreme aerodynamic trajectories in $(a)$ the three-dimensional latent space and $(b)$ its two-dimensional view for the undisturbed baseline cases.
    $(c)$ Undisturbed vorticity fields at $\theta = \pi/4$ and $\pi$ for $\alpha \in [30,60]^\circ$.
    The values inside each snapshot report the level of unsteadiness with $\sigma_{\bm\omega} = ||{\bm\omega}(t)-\overline{\bm\omega}||_2/||\overline{\bm\omega}||_2$.
    }
    % \vspace{-10mm}
    \label{fig5-new}
\end{figure}

With the nonlinear lift-augmented autoencoder, the entire collection of extreme aerodynamic vortical flows spanning over a large parameter space can be compressed into only three latent variables.
The latent vectors ${\bm \xi}(t)$ in the present three-dimensional space are visualized in figure~\ref{fig5-new}$(a)$.
Here, undisturbed baseline cases are shown in color while gray lines correspond to all the trajectories mapped from the disturbed vorticity flow field data.
A variety of vortical flows with and without gust disturbances across five different angles of attack are considered.
All of the extreme aerodynamic cases reside in the vicinity of the undisturbed base states, forming the cone-type structure.
This cone shape is referred to as an inertial manifold to which the long-time dynamics converge~\cite{foias1988modelling,temam1989inertial,de2023data}.
That is, the undisturbed periodic wake dynamics provide the backbone of the manifold with the extreme aerodynamic trajectories lying in the vicinity of this manifold in the three-dimensional latent space.

Here, let us detail the latent trajectories of the undisturbed flows.
The latent vectors for the undisturbed flows across the angle of attack are aligned along the ${\xi}_3$ direction.
% While the different angles of attack can be distinguished in a low-order space, the shape of the trajectory for each angle also reflects the wake behavior of the original high-dimensional space.
The case of $\alpha = 20^\circ$ is mapped as a single dot while the other baseline cases with unsteady periodic shedding at $\alpha \geq 30^\circ$ exhibit cyclic trajectories.
These observations in the latent space correspond to the steady flow at $\alpha = 20^\circ$ and unsteady limit-cycle oscillations at $\alpha \geq 30^\circ$ of vorticity fields.

The two-dimensional (projected) view of the latent space and representative vorticity fields for $\alpha \in [30, 60]^\circ$ at two different phases $\theta = \pi/4$ and $\pi$ are respectively shown in figures~\ref{fig5-new}$(b)$ and $(c)$.
The radius of each limit cycle for the undisturbed cases of $\alpha \geq 30^\circ$ increases with the angle of attack.
They also correlate to the level of unsteadiness present in the vorticity field which we quantify $\sigma_{\bm\omega} = ||{\bm\omega}(t)-\overline{\bm\omega}||_2/||\overline{\bm\omega}||_2$, where $\overline{\bm\omega}$ is a time-averaged vorticity field.
These values are listed in the representative snapshots for $\alpha \in [30, 60]^\circ$ in figure~\ref{fig5-new}$(c)$.
As shown, the value of $\sigma_{\bm\omega}$ increases with the angle of attack.
In other words, the increase of the radius is due to the increase in flow unsteadiness for each angle of attack case.
Furthermore, undisturbed vorticity fields at each phase depicted in figure~\ref{fig5-new}$(c)$ present a similar wake shedding pattern across the angle of attack.
These observations suggest that the undisturbed wakes can be successfully low-dimensionalized while preserving the phase (timing) and amplitude (fluctuation) in the original high-dimensional space.

\begin{figure}
    \centering
    \includegraphics[width=\textwidth]{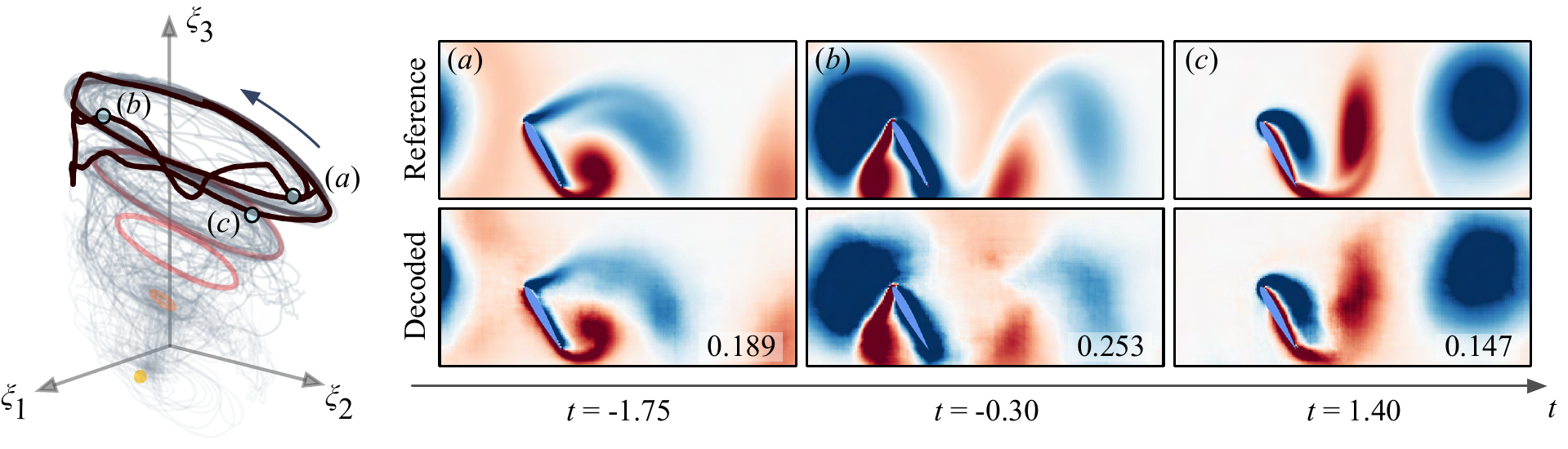}
    \vspace{-2mm}
    \caption{
    Extreme aerodynamic trajectories in the three-dimensional latent space and vortical flow snapshots for $(\alpha, G,D,Y) = (60^\circ,-2.8,1.5,0)$.
    The value inside of each decoded snapshot reports the $L_2$ spatial reconstruction error norm.
    }
    % \vspace{-10mm}
    \label{fig5}
\end{figure}

Next, let us focus on the latent space trajectories for the disturbed wake flows. 
All gray trajectories corresponding to extreme aerodynamic flows reside around the undisturbed orbits.
To investigate the implication of low-dimensionalized extreme aerodynamic trajectories, we take an example case of $(G,D,Y) = (-2.8,1.5,0)$ for which a strong, large vortex gust impinges an airfoil at $\alpha=60^\circ$.
The latent variable trajectory and the reconstructed flow fields over time are also shown in figure~\ref{fig5}.
The value shown in each decoded flow contour reports the spatial $L_2$ reconstruction error norm $\varepsilon = ||{\bm\omega}-\hat{\bm\omega}||_2/||{\bm\omega}||_2$.
The vorticity field can be reconstructed well over time from the three variables with only approximately 20\% error.
This level of error is reasonable for capturing the coherent structures accurately because the spatial $L_2$ norm is a strict comparative measure~\cite{FFT2019a}.
{While this feature of $L_2$ norm is useful for successful training of nonlinear machine-learning models, one can also consider structural similarity index (SSIM) for assessing rotational and translational similarities of vortical flows~\cite{wang2004image,anatharaman2023image}.}
{The error level here is similar across the parameter space.
The lift decoder can also provide accurate estimates of lift coefficient $C_L$, corresponding to approximately 1\% $L_2$ error~\cite{FT2023}.}
This successful reconstruction indicates that the three-dimensional latent variables retain the essence of high-dimensional vortical flows without significant loss of key physics.

The extreme aerodynamic trajectories depicted in figure~\ref{fig5} exhibit the influence of strong vortex gusts on the flow.
From the points $(a)$ to $(b)$ in figure~\ref{fig5}, the latent vector dynamically rises and drops {across the vertical direction in the latent space}.
This is likely because of the approach of negative vortex disturbance to the airfoil, drastically changing the effective angle of attack $\alpha_{\rm eff}$~\cite{anderson1991fundamentals,sedky2020lift,he2020stall}.
In other words, the present lift-augmented autoencoder finds the relationship between extreme aerodynamic flows and lift force in a low-order manner.
While the latent dimension is generally determined by checking the reconstruction performance of autoencoder, we note that the error behavior of the present flows plateaus even if the latent dimension is increased as the flows are well approximated in the three-dimensional space with phase-amplitude ($\theta-r$) coordinates and the effective angle of attack.

The present physically-interpretable low-dimensional representation of extreme aerodynamic flows is obtained due to the lift-augmented network while a regular autoencoder may not necessarily provide an understandable latent data distribution~\cite{FT2023}.
We emphasize that expressing extreme disturbance effects about the undisturbed baseline dynamics is critical in developing flow control strategies because it enables us to identify the desired direction (or control objective) in the low-order coordinates to mitigate the strong impact of extreme vortex gusts.

% \vspace{-6mm}
\section{Phase-amplitude reduction and optimal control}
\label{sec:method}

\begin{figure}
    \centering
    \includegraphics[width=\textwidth]{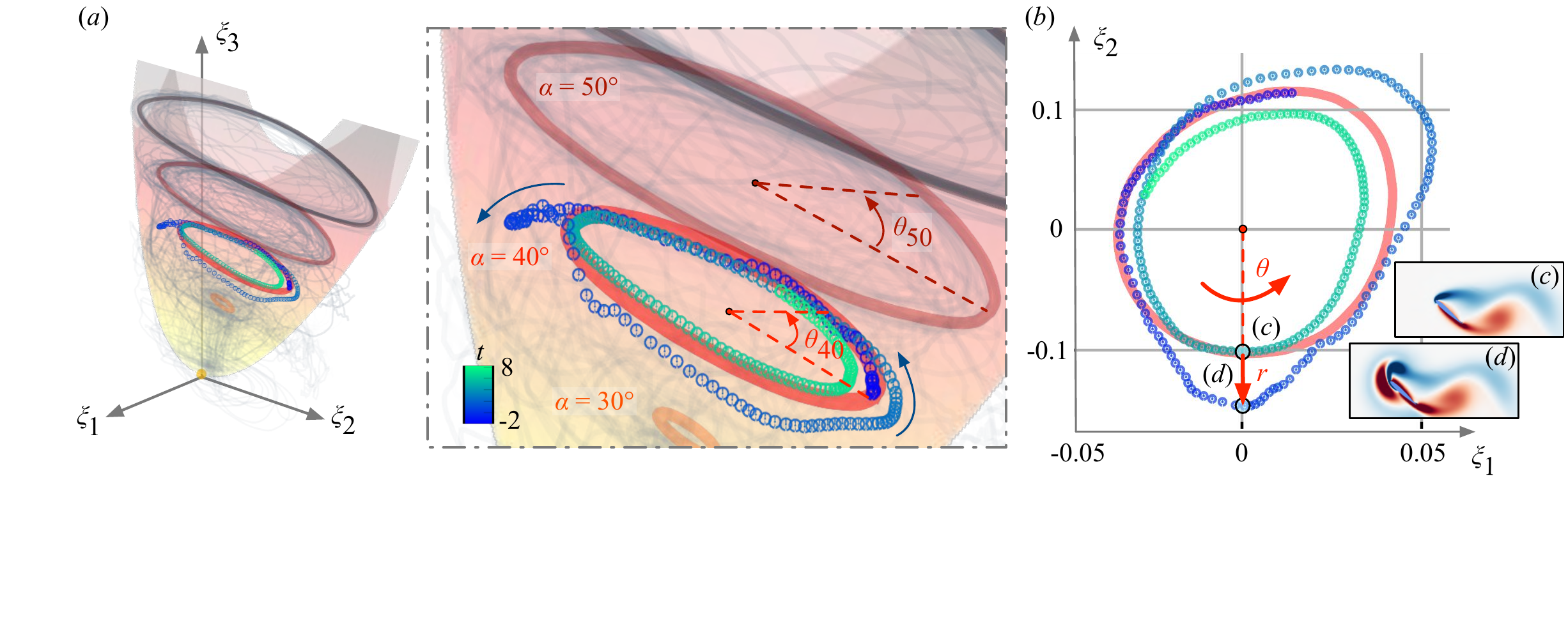}
    \vspace{-5mm}
    \caption{
    $(a)$ The extreme aerodynamic manifold with phase and amplitude.
    The aerodynamic trajectory indicated by the markers, colored by convective time, corresponds to the case of $(\alpha, G,D,Y) = (40^\circ, 2.8,0.5,-0.3)$.
    $(b)$ Two-dimensional plane for $\alpha = 40^\circ$.
    Flow fields at the same phase but different amplitudes chosen from undisturbed and disturbed cases are inserted.
    }
    % \vspace{-10mm}
    \label{fig6}
\end{figure}

With the uncovered latent space representation, we can quantitatively assess the influence of extreme vortex gusts on the dynamics in a low-order manner.
In particular, the present nonlinear coordinate transformation suggests that the vortex-airfoil interaction can be analyzed through the latent space with phase ${\theta}$ and amplitude {deviation} $r$, as illustrated in figure~\ref{fig6}.
The latent variable captures similar wake structures at the same phase $\theta$ while showing the amplitude difference attributed to the vortex-airfoil interaction, as exhibited in figure~\ref{fig6}$(b)$.
This observation suggests that control strategies that push the extreme aerodynamic trajectory towards the direction of the undisturbed baseline state in the latent space mitigate the influence of vortex disturbance in the flow field{, naturally calling for a swift system modification on phase-amplitude coordinates.}

In this study, we analyze and control extreme aerodynamic flows using phase-amplitude modeling with the following three steps:
\begin{enumerate}
    \item dynamical modeling in latent space using {SINDy} (section~\ref{sec:SINDy}),
    \item phase-amplitude reduction to assess phase- and amplitude sensitivity functions (section~\ref{sec:PRA}), and 
    \item control of extreme aerodynamic flows with amplitude-constrained optimal waveform for fast synchronization (section~\ref{sec:AFE}).
\end{enumerate}

Using these steps, we derive a control law to suppress the large fluctuation of lift force due to the vortex disturbance within a very short time duration.
% To derive a control strategy that can quickly modify the flow state while amplitude modulation from the undisturbed baseline dynamics can also be suppressed in the latent space, we consider amplitude-constrained optimal waveform for fast entrainment~\cite{takata2021fast} (section~\ref{sec:AFE}) which can be obtained through phase-amplitude reduction (section~\ref{sec:PRA}) with a dynamical model in a low-order space (section~\ref{sec:SINDy}). 
In this section, we introduce the detailed approach used at each step of the present control strategy.
% , allowing the use of knowledge from the extreme aerodynamic manifold for flow control of strong vortex-airfoil interaction.

\subsection{Sparsity-promoting low-dimensional dynamical modeling}
\label{sec:SINDy}

We model the dynamics of the latent vector ${\bm \xi}$ with a system of ordinary differential equations (ODE) using sparse identification of nonlinear dynamics (SINDy)~\cite{brunton2016discovering}.
This data-driven technique identifies nonlinear model equations from a given time-series data.
Let us consider a dynamical system for the latent vector ${\bm \xi}(t)\in \mathbb{R}^3$,
\begin{align}
    \dot{\bm{\xi}}(t) = \bm{F} ( \bm{\xi}(t) ). \label{eq:state_eq}
\end{align}
The temporally discretized data of $\bm{\xi}$ are collected to prepare a data matrix $\bm{\Xi}$,
\begin{align}
  \bm{\Xi}=
  \left( 
    \begin{array}{c}
      \bm{\xi}^T(t_1)  \\
      \bm{\xi}^T(t_2)  \\
       \vdots  \\
      \bm{\xi}^T(t_m) \\
    \end{array}
     \right)
    =
  \left( 
    \begin{array}{ccc}
      \xi_1(t_1) & \xi_2(t_1) & \xi_3(t_1) \\
      \xi_1(t_2) & \xi_2(t_2) & \xi_3(t_2) \\
       \vdots  &  \vdots &  \vdots \\
      \xi_1(t_m) & \xi_2(t_m) & \xi_3(t_m) \\
    \end{array}
     \right)
     \in \mathbb{R}^{m \times 3},
\end{align}
where $m$ is the number of snapshots.
We also prepare a library matrix $\Phi (\bm{\Xi})$ including nonlinear terms of $\bm{\Xi}$.
This study uses sine and cosine functions for the library matrix construction,
\begin{align}
    \Phi (\bm{\Xi}) = \left( 
      \begin{array}{cccccccc}
         \sin({\bm\Xi}) & \sin({\bm\Xi}/2) & \sin({\bm\Xi}/4) &  \sin({2{\bm\Xi}}) & \sin({4{\bm\Xi}}) 
          & \cdots & \cos({2{\bm\Xi}}) & \cos({4{\bm\Xi}})
      \end{array}
    \right)
    \in \mathbb{R}^{m \times n_l},
    \label{eq:library}
\end{align}
where we include sine and cosine functions of $\xi_i$, $\xi_i/2$, $\xi_i/4$, $2\xi_i$, and $4\xi_i$, resulting in the number of the library series $n_l$ to be 10.
While polynomials constructed by given variables are often considered for the library matrix construction~\cite{brunton2016discovering,BPK2016b,KKB2018,LKLLBK2019}, we have found that a trigonometric function-based library can provide a more accurate solution for the present problem.
{The SINDy-based modeling accuracy is determined by a set of factors including the data set, the number of snapshots, the library functions, and the optimization method~\cite{fukami2020sparse}.}

With the data matrix $\bm{\Xi}$ and the library matrix $\Phi (\bm{\Xi})$, the latent dynamics is modeled as a form of ODE by determining the coefficients for each library term.
A coefficient matrix $\Psi$ is obtained by solving the following regression problem,
\begin{align}
    \dot{\bm{\Xi}}(t)=\Phi (\bm{\Xi}) \Psi, \label{eq:sindy}
\end{align}
with 
\begin{align}
    \Psi = (\psi_{\xi_1} ~~ \psi_{\xi_2}~~ \psi_{\xi_3}) = \left( 
      \begin{array}{ccc}
        \psi _{(\xi_1,~1)} & \psi_{(\xi_2,~1)} & \psi_{(\xi_3,~1)} \\
        \psi _{(\xi_1,~2)} & \psi_{(\xi_2,~2)} & \psi_{(\xi_3,~2)} \\
        \vdots & \vdots & \vdots \\
        \psi _{(\xi_1,~n_l)} & \psi_{(\xi_2,~n_l)} & \psi_{(\xi_3,~n_l)} \\
      \end{array}
    \right).
    \label{eq:coefmat}
\end{align}
In this study, the adaptive Lasso~\cite{Zou2006,fukami2020sparse} is used to optimize the coefficient matrix $\Psi$.
Once we obtain an accurate low-dimensional dynamical model {(equation~\ref{eq:sindy})}, the model is then used to perform the phase-amplitude reduction which provides the optimal timing and location of control actuation to efficiently and quickly modify the dynamics.

\subsection{Phase-amplitude reduction analysis}
\label{sec:PRA}

Here, let us introduce phase-amplitude reduction for a periodic, stable limit-cycle oscillator $\dot{\bm \xi}(t) = {\bm F}({\bm \xi}(t))$ obtained by the SINDy for each angle of attack case.
It is assumed that this ODE has a stable limit-cycle solution ${\bm\xi}_0(t) = {\bm\xi}_0(t+T)$, where $T = 2\pi/\omega_{\alpha}$ with the natural frequency $\omega_{\alpha}$ of the latent variable ${\bm \xi}$ for the undisturbed baseline case at each angle of attack~$\alpha$.
{The natural frequency in the latent dynamics matches that in the high-dimensional wake dynamics as the encoder is applied to a time series of flow snapshots.}

Given the aforementioned ODE system for the undisturbed system at each angle of attack, we define the phase and amplitude variables $\theta$ and $r$ of the latent system state ${\bm \xi}$, as illustrated in {figures~\ref{fig5-new} and~\ref{fig6}},
\begin{align}
    \theta = \Theta (\bm \xi), ~~~ r = R(\bm \xi),
\end{align}
where ${\Theta}({\bm \xi})$ and $R({\bm \xi})$ are the phase and amplitude functions, respectively.
Here, the phase and amplitude functions provide a global linearization of the original nonlinear dynamics in the basin of attraction of the limit cycle for each angle of attack.
% Given the aforementioned ODE system, the phase and amplitude in the basin of $\chi$ can be expressed with ${\Theta}({\bm \xi})$ and $R_i({\bm \xi})$.
The phase function $\Theta$ is defined to satisfy the condition that the phase $\theta$ increases with a frequency $\omega_\alpha$ at angle of attack $\alpha$.
Hence, the generalized phase dynamics is described as
\begin{align}
    \dot{\Theta}({\bm \xi}) 
    = \langle \nabla{\Theta}({\bm \xi}), \dot{\bm \xi} \rangle 
    = \langle \nabla{\Theta}({\bm \xi}), {\bm F}({\bm \xi}) \rangle 
    = \omega_{\alpha},
    % &\dot{R}_i({\bm \xi}) 
    % = \langle \nabla{R_i}({\bm \xi}),  \dot{\bm \xi} \rangle 
    % = \langle \nabla{R_i}({\bm \xi}), {\bm F}({\bm \xi}) \rangle 
    % = \lambda_i R_i({\bm\xi}),
\end{align}
% for the latent variable ${\bm \xi}$. 
where $\langle{\bm a},{\bm b}\rangle = \sum_{i=1}^N a_i^*b_i$ is a scalar product.

Similarly, the generalized amplitude dynamics can also be derived with the assumption that $r$ exponentially decays to zero as ${\dot r}=\lambda r$ of the limit cycle, as presented in figure~\ref{fig6}.
Here, $\lambda$ denotes the decay rate given by the Floquet exponent, which characterizes the linear stability of ${\bm\xi}_0$.
The connections among the natural frequency $\omega_\alpha$, phase sensitivity function~$\Theta$, Floquet exponents, and amplitude sensitivity function $R$ ease the evaluation of phase and amplitude functions as discussed later.
Although there are generally $n-1$ amplitudes for $n$-dimensional oscillators associated with the Floquet exponents $\lambda_i$, the dominant, slowest-decaying dynamics is sought whose exponent is denoted as $\lambda$ for simplicity~\cite{nakao2021phase}.
Thus, the amplitude function needs to satisfy
\begin{align}
    \dot{R}({\bm \xi}) 
    = \langle \nabla{R}({\bm \xi}),  \dot{\bm \xi} \rangle 
    = \langle \nabla{R}({\bm \xi}), {\bm F}({\bm \xi}) \rangle 
    = \lambda R({\bm\xi}).
\end{align}

Now considering an external control input ${\bm f}(t)$ to the system, the oscillator dynamics is described as
\begin{align}
    \dot{\bm \xi}(t) = {\bm F}({\bm \xi}(t)) + {\bm f}(t).
\end{align}
For this perturbed system, the dynamics of phase $\theta$ and amplitude $r$ satisfy
\begin{align}
    \dot{\theta}(t) = \omega_{\alpha} +  \langle \nabla\Theta({\bm \xi}(t)), {\bm f}(t)  \rangle, \nonumber\\
    \dot{r}(t) = \lambda r(t) + \langle \nabla R ({\bm \xi}(t)), {\bm f}(t)  \rangle.
\end{align}
Here, we further assume that the control input ${\bm f}(t)$ is of ${\cal O}(\epsilon)$ with $0 < \epsilon \ll 1$.
These equations can be then approximated by neglecting the terms of order ${\cal O}(\epsilon^2)$,
\begin{align}
    \dot{\theta} = \omega_{\bm \alpha} + \langle {\bm Z}(\theta),{\bm f}(t) \rangle, ~~~ {\dot r} = \lambda r + \langle {\bm Y}(\theta),{\bm f}(t) \rangle,
\end{align}
where ${\bm Z}(\theta)=\nabla \Theta|_{{\bm \xi}={\bm\xi}_0(\theta)}$ and ${\bm Y}(\theta)=\nabla {R}|_{{\bm \xi}={\bm\xi}_0(\theta)}$ are the phase and amplitude sensitivity functions, respectively, evaluated on the limit cycle for each angle of attack $\alpha$.

The phase sensitivity function ${\bm Z}(\theta)$ describes the sensitivity of the system phase and the amplitude sensitivity function ${\bm Y}(\theta)$ reveals the sensitivity of the system amplitude about the periodic orbit against external forcing.
Although it is generally difficult to measure the phase and amplitude sensitivity functions, they can be obtained by assessing the left Floquet eigenvectors if a dynamical model is explicitly given~\cite{kuramoto1984chemical,takata2021fast}.

If a low-order model is available through SINDy, we can derive from Floquet theory the phase and amplitude sensitivity functions ${\bm Z}(\theta)$ and ${\bm Y}(\theta)$, respectively.
Here, we introduce the right and left Floquet eigenvectors ${\bm U}_i$ and ${\bm V}_i$ that are the $T$-periodic solutions, 
\begin{align}
    &{\dot{\bm U}_i}(t) = [{\cal J}({\bm \xi}_0(t)) - \lambda_i]{\bm U}_i(t),\nonumber\\
    &{\dot{\bm V}_i}(t) = -[{\cal J}({\bm \xi}_0(t))^{\dag} - \lambda_i^\dag]{\bm V}_i(t),~\label{eq:adj}
\end{align}
{for $i=0,1,..., N-1$,} where the superscript $\dag$ represents the Hermitian conjugate and ${\cal J}$ is a $T$-periodic Jacobian matrix of $\bm F$ evaluated about ${\bm \xi}={\bm \xi}_0(t)$~\cite{ermentrout1996type,brown2004phase,shirasaka2017phase,kuramoto2019concept}.
The phase sensitivity function ${\bm Z}(\theta)$ and the dominant amplitude sensitivity function ${\bm Y}(\theta)$, respectively, are then expressed as
\begin{align}
    {\bm Z}(\theta) = {\bm V}_0(\theta/\omega_{\bm \alpha}), ~~~ {\bm Y}(\theta) = {\bm V}_1(\theta/\omega_{\bm \alpha})~\label{eq:PSF-ASF}
\end{align}
for $0 \leq \theta < 2\pi$.
{To obtain the phase and amplitude sensitivity functions ${\bm Z}(\theta)$ and ${\bm Y}(\theta)$, we first solve the ODE in the forward direction (i.e., the direct problem).
The adjoint equation is then solved once the Jacobian at each phase for the time period is available so that ${\bm U}_1$ and ${\bm V}_1$ can be calculated~\cite{takata2021fast}.}

\subsection{Optimal fast flow control with amplitude constraint}
\label{sec:AFE}

Next, we consider feedforward control based on the phase and amplitude sensitivity functions.
As illustrated in figure~\ref{fig6}, suppressing the amplitude modulation in the low-order space can lead to the mitigation of the gust impact.
Furthermore, {since we now have a clear direction in the phase-amplitude space to mitigate the impact of gusts}, it is possible to achieve fast synchronization with amplitude penalty such that the latent dynamics quickly returns to the undisturbed baseline dynamics while suppressing amplitude deviations~\cite{harada2010optimal,zlotnik2016phase,takata2021fast}.
For vortex-airfoil interaction, synchronization at a higher frequency than the natural frequency with amplitude constraints would provide smaller vortical structures in a wake that are weaker than the undisturbed baseline case~\cite{zhang2022wake,godavarthi2023optimal}, thereby swiftly reducing the vortex gust impact in the high-dimensional space.
Hence, the objective of the present controller is to quickly attenuate the transient dynamics in the low-dimensional latent space ${\bm \xi}$ with phase locking.
We obtain the actuation pattern to achieve the above objective by leveraging the optimal-synchronization waveform with amplitude suppression~\cite{takata2021fast}.

To begin with, let us introduce the relative phase (phase difference) $\phi(t) = \theta(t) - \omega_{\bm f}t$ where $\omega_{\bm f}$ is the forcing signal frequency.
Assuming that the the control input ${\bm f}$ is given in the form of ${\bm f}(t) = {\bm b}_{\bm \xi}(\omega_{\bm f} t)$, the phase dynamics becomes
\begin{align}
    \dot{\theta} = {\omega}_{\bm \alpha} +  \langle {\bm Z}(\theta),{\bm b}_{\bm \xi}(\omega_{\bm f} t) \rangle.
\end{align}
The dynamics of the relative phase is provided as
\begin{align}
    \dot{\phi}(t) 
    &= \Delta \Omega +  
    % \dfrac{1}{T_{\bm f}} \int_0^{T_{\bm f}} 
    \langle {\bm Z}(\phi + \omega_{\bm f} t),{\bm b}_{\bm \xi}(\omega_{\bm f} t) \rangle,
    % d\tau
\end{align}
where $T_{\bm f}$ is a period of the periodic forcing input and $\Delta \Omega = {\omega}_{\alpha}-\omega_{\bm f}$.
Since this equation is non-autonomous, we consider deriving an autonomous form by averaging over a period of forcing~\cite{kuramoto1984chemical,hoppensteadt1997weakly}.
The asymptotic behavior of the relative phase dynamics can be approximated as 
\begin{align}
    {\dot\phi}(t) = \Delta \Omega + \Gamma(\phi), ~~~ \Gamma(\phi) =\dfrac{1}{T_f}\int_0^{T_{\bm f}} 
    \langle {\bm Z}(\phi + \omega_{\bm f} \tau),{\bm b}_{\bm \xi}(\omega_{\bm f} \tau) \rangle d\tau, 
\end{align}
where $\Gamma(\phi)$ is called the phase coupling function.
Phase locking can be achieved if the relative phase becomes a constant such that $\dot\phi \rightarrow 0$.
This phase locking is achieved when $- \max\Gamma(\phi) < \Delta \Omega < - \min\Gamma(\phi)$, uncovering the Arnold tongue that captures the condition for synchronization~\cite{shim2007synchronized}.

Next, we seek the optimal input to achieve the present control objective.
The controller is first asked to synchronize the system to a forcing (target) frequency as quickly as possible.
In other words, the rate of convergence of $\phi$ to a fixed, stable phase-locking point $\phi^*$ needs to be maximized to satisfy $\dot{\phi^*} = \Delta \Omega + \Gamma (\phi^*) = 0$.
Furthermore, we also aim to suppress the excitation from the limit-cycle dynamics in the latent space.

To derive the periodic waveform that can satisfy the above conditions, the following cost function is used to formulate an optimization problem,
\begin{align}
    {\cal L}({\bm b}_{\bm \xi}) = 
    &- \Gamma^\prime (\phi^*) 
    + \nu \biggl(P - \dfrac{1}{T_f}\int_0^{T_f}
    \langle {\bm b}_{\bm \xi}(\omega_f\tau) \cdot {\bm b}_{\bm \xi}(\omega_f\tau) \rangle d\tau\biggr) \nonumber\\
    &+ \mu (\Delta \Omega + \Gamma (\phi^*)) 
    - 
    % \sum_{i=1}^{M}
    k \biggl(\dfrac{1}{T_f}\int_0^{T_f}
    |\langle {\bm Y} (\phi^*+\omega_{\bm f} \tau), {\bm b}_{\bm \xi}(\omega_{\bm f} \tau)\rangle|^2 d\tau\biggr),
    \label{eq:cost}
\end{align}
where 
% $i=1,...,M$ with the first $M$ amplitude variables,
$\nu$ and $\mu$ are Lagrangian multipliers, and $P$ is a constant satisfying $\sqrt{P} \sim {\cal O}(\omega_{\bm \xi}\delta)$.
The first term contributes to maximizing the synchronization rate $-\Gamma^\prime (\phi^*)$, the second term constrains the energy of actuation, and the third term directly corresponds to the rate of convergence of $\phi$.
In addition, the fourth term penalizes the excitation of the amplitude variable of amplitude sensitivity function with the weight $k$.

The above optimization problem can be solved using the calculus of variations~\cite{zlotnik2013optimal,takata2021fast} once we obtain the phase and amplitude sensitivity functions ${\bm Z}(\theta)$ and ${\bm Y}(\theta)$, respectively, through Floquet analysis for the latent evolution equation derived by SINDy.
We can finally derive the optimal waveform as 
\begin{align}
    {\bm b}_{\bm \xi}(\omega_{\bm f} t) =
    \dfrac{1}{2} 
    \biggl[ \nu {\bm I} + 
    % \sum_{i=1}^M 
    k
    {\rm Re}({\bm Y}(\phi^* + \omega_{\bm f} t) {{\bm Y}^*}^\dag (\phi^* + \omega_{\bm f} t)) 
    \biggr]^{-1}
    \biggl[-{\bm Z}^\prime (\phi^* + \omega_{\bm f} t) + \mu {\bm Z}(\phi^* + \omega_{\bm f} t)\biggr],~\label{eq:wave}
\end{align}
where 
% $i=1,...,M$ and 
${\bm Z}^\prime$ is {the derivative of phase sensitivity function with respect to phase} and ${\bm I}$ is an identity matrix.
{The weight value $k$ can be chosen either empirically or through the $L$-curve analysis~\cite{l-curve} to balance the terms for fast synchronization and the amplitude constraint~\cite{takata2021fast}.}

\begin{figure}
    \centering
    \includegraphics[width=0.9\textwidth]{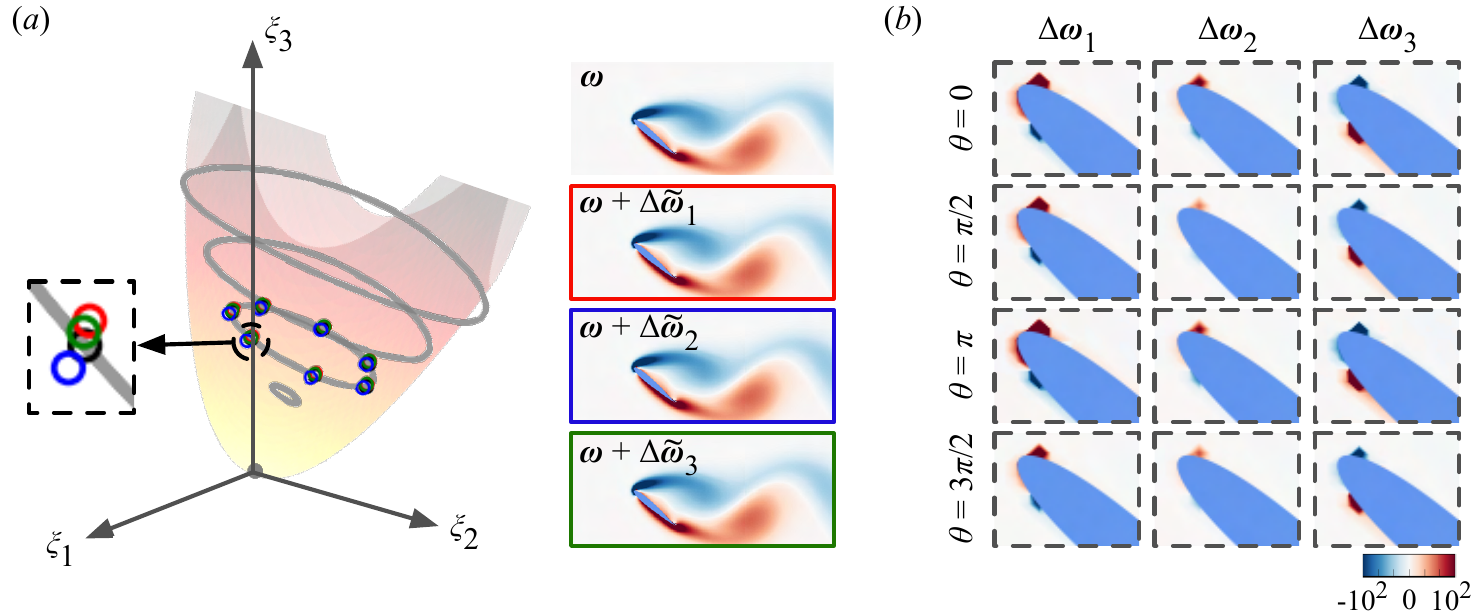}
    \vspace{-2mm}
    \caption{
    Conversion from latent perturbation to forcing in the original space.
    $(a)$~Examples of perturbed vorticity fields ${\bm \omega} + \Delta \omega$ and the corresponding latent vectors ${\bm \xi} + \Delta \tilde{\bm \xi}$.
    The color used for the points in the latent space corresponds to the flame color for the vorticity field. 
    $(b)$ The perturbation in the high-dimensional space toward a particular direction~$\Delta {\bm \omega}_i$.
    }
    \vspace{-2mm}
    \label{fig7}
\end{figure}

Because the optimal waveform in equation~\ref{eq:wave} is derived in the latent space, we need to convert it to forcing in the original physical space.
Here, we derive the relationship of the perturbation between the latent and physical spaces,  $\Delta{\bm \xi}$ and $\Delta{\bm \omega}$, respectively, for which we assume that the encoder ${\cal F}_e$ is continuously differentiable.
To find such a perturbation $\Delta{\bm \omega}$ toward a particular direction in the physical space, we consider an input vorticity field ${\bm \omega}(t^*)$ with an arbitrary perturbation in the high-dimensional space ${\Delta \tilde{\bm \omega}(t^*)}$.
The latent vector corresponding to the given vorticity field can be approximated with the Jacobian matrix ${\cal J}_{\bm \xi}(\bm \omega)$ of ${\cal F}_e$ evaluated at time $t^*$ such that
\begin{align}
    {\bm \xi} + \Delta \tilde{\bm \xi} &= {\cal F}_e ({\bm \omega}+\Delta \tilde{\bm \omega})\nonumber\\
    & \simeq {\cal F}_e({\bm \omega}) + {\cal J}_{\bm \xi}(\bm \omega) \Delta \tilde{\bm \omega}.~\label{eq:jacob}
\end{align}
For the current three-dimensional latent-vector model, we consider giving three different patterns of perturbation to the physical flow field through $\Delta \tilde{\bm \omega}_m$ $(m=1,2,3)$.
From equation~\ref{eq:jacob}, the deviation of the latent vector due to the perturbation in the physical space is expressed as
\begin{align}
    % \Delta {\bm \xi}_A = {\cal J}_{\bm \xi}\Delta {\bm q}_A, ~~~ \Delta {\bm \xi}_B = {\cal J}_{\bm \xi}\Delta {\bm q}_B, ~~~\Delta {\bm \xi}_C = {\cal J}_{\bm \xi}\Delta {\bm q}_C,~\label{eq:jacobdel}
    \Delta \tilde{\bm \xi}_m = {\cal J}_{\bm \xi}\Delta \tilde{\bm \omega}_m.~\label{eq:jacobdel}
\end{align}
For the unit vectors ${\bm e}_1$, ${\bm e}_2$, and ${\bm e}_3$ in the latent space, the relationship between the perturbation in each direction of the latent dynamics and the deviations in equation~\ref{eq:jacobdel} is expressed through a coefficient matrix ${\bm H} \in \mathbb{R}^{3\times 3}$,
\begin{align}
    {\bm I} = 
    \left( 
    \begin{array}{ccc}
      % | & | & | \\
      {\bm e}_1 & {\bm e}_2 & {\bm e}_3 %\\
      % | & | & | \\
    \end{array}
     \right)
    &=
  \left( 
    \begin{array}{ccc}
      % | & | & | \\
      \Delta \tilde{\bm \xi}_1 & \Delta \tilde{\bm \xi}_2  & \Delta \tilde{\bm \xi}_3 %\\
      % | & | & | \\
    \end{array}
     \right)
  \left( 
    \begin{array}{ccc}
      H_{11} & H_{12} & H_{13} \\
      H_{21} & H_{22} & H_{23} \\
      H_{31} & H_{32} & H_{33} \\
    \end{array}
     \right)\nonumber\\
    &= {\bm D}_{\bm \xi} {\bm H}.
\end{align}
Hence, to individually perturb the latent system in the ${\bm e}_1$, ${\bm e}_2$, and ${\bm e}_3$ directions, the perturbation in the high-dimensional space toward a particular direction $\Delta {\bm \omega}_i$ $(i=1,2,3)$ is derived as
\begin{align}
    % \Delta {\bm q}_1 = {H}_{11}\Delta{\bm q}_A + {H}_{21}\Delta{\bm q}_B + {H}_{31}\Delta{\bm q}_C,\nonumber\\
    % \Delta {\bm q}_2 = {H}_{12}\Delta{\bm q}_A + {H}_{22}\Delta{\bm q}_B + {H}_{32}\Delta{\bm q}_C,\nonumber\\
    % \Delta {\bm q}_3 = {H}_{13}\Delta{\bm q}_A + {H}_{23}\Delta{\bm q}_B + {H}_{33}\Delta{\bm q}_C,
    % \label{eq:relation}
    \Delta {\bm \omega}_i =  \sum_{j=1}^3 H_{ji} \Delta \tilde{\bm \omega}_j,
    \label{eq:relation}
\end{align}
where the coefficient matrix can be determined as ${\bm H} = {\bm D}^{-1}_{\bm \xi}$.

In the present study, the three different perturbations in the physical space (for equation~\ref{eq:jacobdel}) are determined by a momentum injection at the leading edge of the airfoil at $45^\circ$, $90^\circ$, and $135^\circ$ relative to the local tangential direction.
{The actuation cost with the steady momentum coefficient $c_\mu = (\rho u_{\rm jet}^2 \sigma)/(0.5 \rho u_\infty^2 c)$, where $u_{\rm jet}$ is the actuation velocity and $\sigma$ is the actuator width,} is set to be 0.016.
Here, three perturbations in equation~\ref{eq:jacobdel} are individually derived at each $\theta$ over the periodic dynamics because the linearization in equation~\ref{eq:jacob} is locally valid for small perturbation.

The perturbed flow fields and latent vectors (in equation~\ref{eq:jacob}) and the derived forcing in the high-dimensional space corresponding to a perturbation for each direction in the latent space (in equation~\ref{eq:relation}) are shown in figure~\ref{fig7}.
{The shift of latent vector shown as red, blue, and green circles at each phase in figure~\ref{fig7}$(a)$ is quite small due to the small forcing input of $\Delta \tilde{\bm \omega}$.
The magnitude and shape of forcing structures depicted in figure~\ref{fig7}$(b)$ vary over the dynamics and across the latent variables.
The designed forcing is localized due to small forcing inputs for the three different patterns of $\Delta {\bm \omega}$.}
In this study, we examine how extreme aerodynamic flows can be controlled by such localized actuation spanning over a very small area with the assistance of optimal waveform analysis within a very short time duration.

The identified relationship of the perturbation in the latent and high-dimensional spaces is also used to verify the phase sensitivity function ${\bm Z}(\theta)$ which is derived in equation~\ref{eq:PSF-ASF} (details to be provided later {with figure~\ref{fig10}}).
Here, the designed perturbation ${\bm K}_\omega({\bm x},t)= \epsilon {\bm b}_{\bm \xi}(t)\Delta {\bm \omega}({\bm x},t)$ is added to the right-hand side of the vorticity transport equation as 
\begin{align}
    \partial_t {\bm \omega}({\bm x},t) = -{\bm u}\cdot \nabla {\bm \omega} 
    % + {\bm q}\cdot \nabla {\bm u}
    +{Re}^{-1} \nabla^2 {\bm \omega} + {\bm K}_{\bm \omega}({\bm x},t),
\end{align}
where the designed perturbation in the velocity form ${\bm K}_{\bm u} = \nabla \times {\bm K}_\omega$ is used for performing the present simulations~\cite{kawamura2022adjoint}.
In the next section, we assess the amount of attenuation that can be achieved for extreme aerodynamic flows with the present localized forcing and the optimal waveform.

\section{Results and discussion}
\label{sec:res_dis}

Let us present the data-driven and phase-amplitude-inspired modeling and control of extreme aerodynamic flows.
We consider strong vortex-airfoil interactions at $\alpha=40^\circ$ as examples.
The validity of a low-dimensional dynamical model identified by SINDy is first considered.
Once the phase and amplitude sensitivity functions are evaluated with Floquet analysis of the identified low-order model, we apply the present control strategy to the extreme aerodynamic flows for gust mitigation.

\subsection{Identification of low-dimensional latent dynamics}
\label{sec:SINDyresult}

\begin{figure}
    \centering
    \includegraphics[width=1\textwidth]{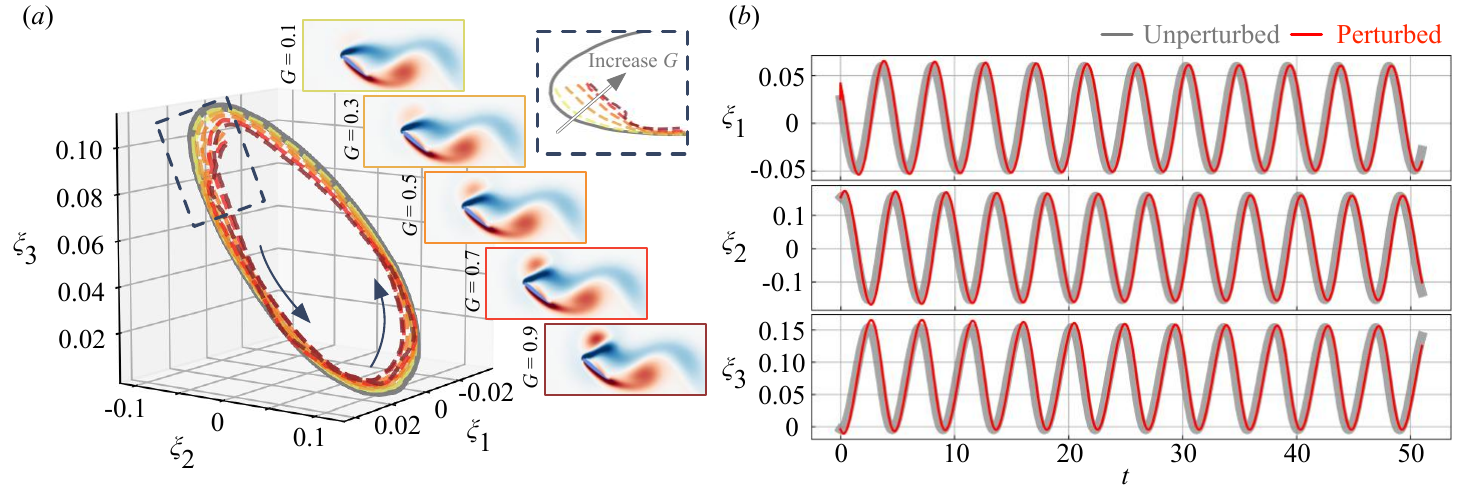}
    \vspace{-4mm}
    \caption{
    $(a)$ Weakly-disturbed transient data used for SINDy training.
    The latent variables and the initial vorticity snapshot for cases with a positive vortex gust with $Y=0.1$ are visualized.
    A zoomed-in view of the latent space is also shown.
    $(b)$~SINDy-based latent dynamics identification.
    {Unperturbed and perturbed model dynamics at $t=0$ are shown.}
    }
    % \vspace{-10mm}
    \label{fig8}
\end{figure}

Here, we discuss the SINDy-based low-dimensional latent dynamical modeling.
As presented in section~\ref{sec:SINDy}, SINDy requires a data matrix $\bm{\Xi}$ and its time derivative $\dot{\bm{\Xi}}$ to approximate the dynamics through regression.
To accurately model the dynamics to measure both the phase and amplitude sensitivity functions, giving only the data of the undisturbed periodic oscillation is insufficient because the identified model does not incorporate dynamics of the limit cycle.
{In other words, the low-order model needs to learn how the dynamics return to the baseline orbit when giving a perturbation~\cite{YFTN2024}.}
% The SINDy-based modeling with periodical latent vectors generally finds the energy-conserving Hamiltonian system which cannot cover dissipative nature appearing in a high-dimensional flow physics~\cite{fukami2020sparse}.
% This indicates that the time trajectory of the Hamiltonian system does not return to the original limit cycle after the perturbation and would create a new periodic oscillation with a different radius.
% Since phase-amplitude reduction analysis assumes a $T$-periodic, stable limit cycle function, correctly identifying the non-Hamiltonian dynamics in a low-order manner is critical.
For this reason, the present training data for SINDy includes not only the periodic oscillation but also the transient process of weakly disturbed cases with a vortex gust.
Examples of training vorticity snapshots and corresponding latent vectors are shown in figure~\ref{fig8}$(a)$.
To consider the transient process, the latent vector ${\bm \xi}$ and their time derivatives $\dot{\bm \xi}$ corresponding to twenty cases with a parameter combination of $G = (\pm 0.1, \pm 0.3, \pm 0.5, \pm 0.7, \pm 0.9)$, $Y= (-0.3, 0.1)$, and $D=0.5$ are prepared.
To accurately learn how the dynamics return to the periodic limit cycle, the snapshots after the vortex disturbance passes over an airfoil are used.

{To assess if the model learns the vicinity of limit-cycle dynamics, the model is integrated with perturbations added at $t=0$, as shown in figure~\ref{fig8}$(b)$.}
After the perturbation at $t=0$, the amplitude gradually returns to the original level across all latent vectors. 
This reflects the given airfoil wake physics in the high-dimensional space in which the effect of perturbation dies out over the convection and the wake dynamics come back to the undisturbed periodic shedding oscillation{, which is critical to accurately perform phase-amplitude reduction analysis in a low-order space.}

\subsection{Phase-amplitude-based modeling of latent dynamics}
\label{sec:PAresult}

Given the identified model equation, we can apply the phase-amplitude model reduction to obtain the phase and amplitude sensitivity functions.
As expressed in equations~\ref{eq:adj} and~\ref{eq:PSF-ASF}, these functions can be obtained by assessing the left Floquet eigenvectors.
These two functions for the present latent dynamics over $0\leq \theta < 2\pi$ are depicted in figures~\ref{fig10}$(a)$ and $(b)$.

The phase sensitivity functions in the $\xi_1$ and $\xi_2$ directions are much greater in magnitude than that for $\xi_3$.
This is because the latent variables $\xi_1$ and $\xi_2$, which mainly compose the phase plane as illustrated in figure~\ref{fig6}, possess a larger variation over the dynamics compared to $\xi_3$ capturing the effective angle of attack on the present manifold.
This indicates that perturbing the system in the $\xi_1$ and $\xi_2$ directions is effective in modifying the dynamics from the aspect of phase delay or advancement.

On the other hand, the relative magnitude of the amplitude sensitivity function ${\bm Y}(\theta)$ for $\xi_3$ is of a similar order to that for the other two variables.
This implies that the perturbation in the $\xi_3$ direction can contribute to the amplitude modulation of the latent dynamics.
This also agrees with the aerodynamic insight in a high-dimensional space in which pitching the wing (in the $\xi_3$ direction) greatly modifies the fluctuation from the mean state of periodic wake shedding.

While being able to derive the phase and amplitude sensitivity functions, these model-based sensitivity functions can be verified by perturbing the vorticity field in a numerical simulation via the conversion, equation~\ref{eq:relation}, and directly assessing the phase shift over the dynamics.
The phase sensitivity functions evaluated in this manner are also plotted with circles in figure~\ref{fig10}$(a)$.
The verified results with the forcing are in excellent agreement with the model-based phase sensitivity function, indicating that the SINDy-based model successfully captures the asymptotic flow behavior.
Note that the amplitude sensitivity function ${\bm Y}(\theta)$ is not compared to the one measured in the simulation because it is challenging to directly measure ${\bm Y}(\theta)$ in the high-dimensional system~\cite{nakao2021phase}.

\begin{figure}
    \centering
    \includegraphics[width=0.6\textwidth]{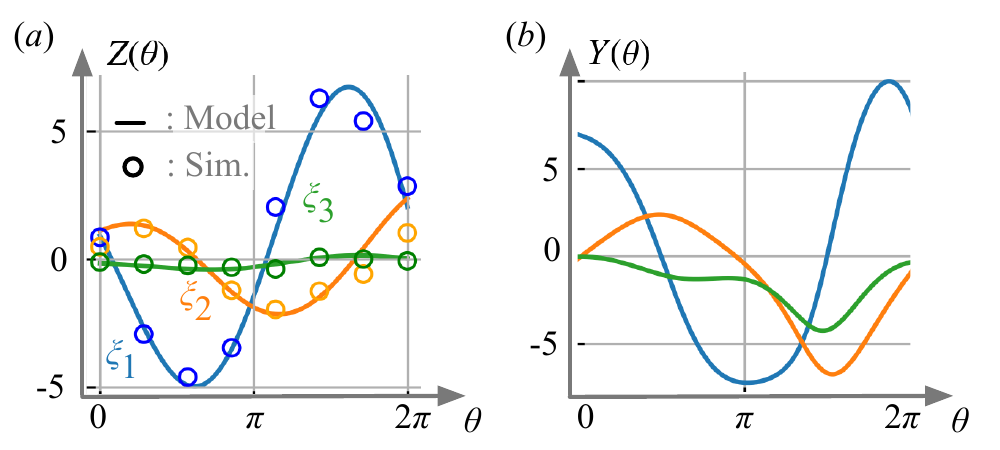}
    % \vspace{-2mm}
    \caption{
    $(a)$ Phase sensitivity function ${\bm Z}(\theta)$ and $(b)$ amplitude sensitivity function ${\bm Y}(\theta)$ for the latent vector $\bm \xi$.
    For ${\bm Z}(\theta)$, the analytical result through the Floquet analysis ($-$: Model) and the verified result with the forcing in equation~\ref{eq:relation} ($\circ$: {Simulation}) are shown.
    }
    % \vspace{-10mm}
    \label{fig10}
\end{figure}

\subsection{Amplitude-constrained fast synchronization control}
\label{sec:cont_result}

With the phase and amplitude sensitivity functions in hand, we are ready to derive the optimal fast synchronization waveform through equations~\ref{eq:cost} and~\ref{eq:wave}.
In this section, we demonstrate how the present method can mitigate the gust impact within a very short time by adding the forcing shown in figure~\ref{fig7} with the optimal waveform.

Let us first apply the present control to the case of $(\alpha, G,D,Y)=(40^\circ, 2.8, 0.5, -0.3)$ to examine the applicability for which a strong counter-clockwise vortex impinges a wing.
The lift coefficient in this case is violently affected due to the approach of the extreme vortex gust and shows significant fluctuation over a short time of less than 1 convective time.
Our aim is to suppress such a sharp force fluctuation within a short time.
We initiate actuation at $t=-1.58$ when a vortex gust appears at the left edge of the domain of the interest.
% This timing is quantitatively determined based on the difference between the pressure sensor measurements under undisturbed flight and a scenario with vortex gusts, as detailed in the Appendix.

To quantify the control effect, we consider the percentage change of lift fluctuation, 
\begin{align}
    \eta = (\Delta C_{L,{\rm ctrl}} - \Delta C_{L,{\rm noctrl}})/\Delta C_{L,{\rm noctrl}},
\end{align}
where $\Delta C_L \equiv \max (C_L) - \min (C_L)$ over $-1.58<t<2$ (during a vortex gust impinges a wing) with the subscripts $(\cdot)_{\rm noctrl}$ and $(\cdot)_{\rm ctrl}$ being uncontrolled and controlled variables, respectively.
Hence, a negative $\eta$ corresponds to suppression of the lift fluctuation.

To derive the optimal waveform through equation~\ref{eq:wave}, the ratio between the natural frequency $\omega_{\alpha}$ and the target frequency $\omega_{\bm f}$, $\omega_{\bm f}/\omega_{\alpha}$, is set to be 1.5 in this case.
The choice of target frequency $\omega_{\bm f}$ is motivated to quickly modify a flow state since the actuation with $\omega_{\bm f}/\omega_{\alpha}>1$ provides faster flow modification than that with $\omega_{\bm f}/\omega_{\alpha}<1$~\cite{godavarthi2023optimal}.
For vortex-gust airfoil interaction, it is anticipated that the impact of the gust can be mitigated quickly by changing the vortex-shedding frequency while suppressing the lift excitation due to the vortex gust.
Hereafter, we consider the waveform and forcing derived by the latent variable $\xi_3$ based on our aerodynamic knowledge that the effective angle of attack captured by ${\xi_3}$ strongly relates to the lift coefficient.
% corresponding to the effective angle of attack.
% While the amplitude sensitivity function ${\bm Y}(\theta)$ for $\xi_1$ and $\xi_2$ possesses a similar order to that for ${\xi}_3$, we have confirmed that the actuation designed with $\xi_1$ and $\xi_2$ cannot suppress the lift fluctuation.
% This is likely because the suppression of amplitude modulation for $\xi_1$ and $\xi_2$ in the latent space does not correspond to that of the lift fluctuation occurring in the high-dimensional space.
% In contrast, we can expect that the actuation based on ${\xi_3}$ would work better based on our aerodynamic knowledge since the effective angle of attack captured by ${\xi_3}$ strongly relates to the lift coefficient.
This study chooses an actuation amplitude of $\epsilon = 0.12$, corresponding to $c_\mu=0.24$, to achieve entrainment for extreme aerodynamic flows while ensuring that the actuated dynamics are under the valid regime of phase-amplitude reduction~\cite{godavarthi2023optimal}.

\begin{figure}
    \centering
    \includegraphics[width=\textwidth]{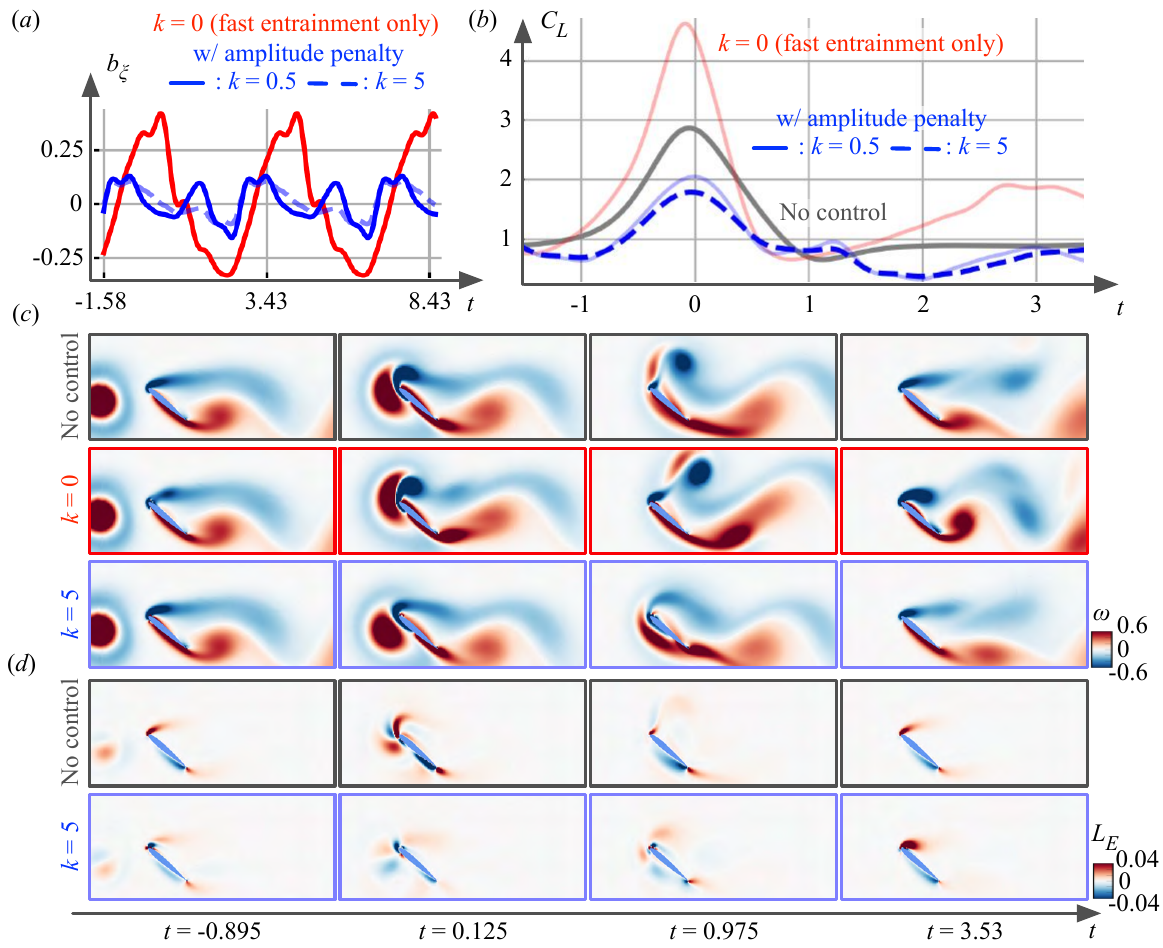}
    \vspace{-2mm}
    \caption{
    Phase-amplitude-based control of an extreme aerodynamic flow of $(\alpha, G,D,Y)=(40^\circ, 2.8, 0.5, -0.3)$.
    $(a)$ Optimal waveform ${b}_{\xi}$ with $k=0$, 0.5, and 5.
    $(b)$ Lift coefficient $C_L$ of the uncontrolled and controlled cases with $k=0$, 0.5, and 5.
    $(c)$ Vorticity fields and $(d)$ lift force elements of the uncontrolled and controlled cases with $k=0$ and 5.
    }
    % \vspace{-10mm}
    \label{fig11}
\end{figure}

The optimal waveform for the case of $(\alpha, G,D,Y)=(40^\circ, 2.8, 0.5, -0.3)$ is shown in figure~\ref{fig11}$(a)$.
To examine the effect of amplitude penalty constrained via equation~\ref{eq:cost}, we consider three different weights, namely $k=0, 0.5$, and 5.
The waveform with amplitude penalty provides a more deformed pattern compared to that with $k=0$ designed for purely fast synchronization only, analogous to the observation with several low-dimensional ODE models by Takata et al.~\cite{takata2021fast}.

This wave pattern with amplitude penalty provides enhanced suppression of the transient lift fluctuation.
The time series of the lift coefficient for each case is presented in figure~\ref{fig11}$(b)$.
While the lift coefficient with the waveform with $k=0$ is more amplified, the actuation with the amplitude-constrained optimal waveform successfully suppresses the lift fluctuation, achieving $\eta = -0.357$.
We emphasize that the present optimal flow modification strategy is designed with minimal computational cost since all procedures expressed in section~\ref{sec:method} are performed in the three-dimensional latent space.
% This suggests that the present method exhibits the potential to significantly and quickly reduce the impact of the extreme vortex gust on a wing in real time.

\begin{figure}
    \centering
    \includegraphics[width=\textwidth]{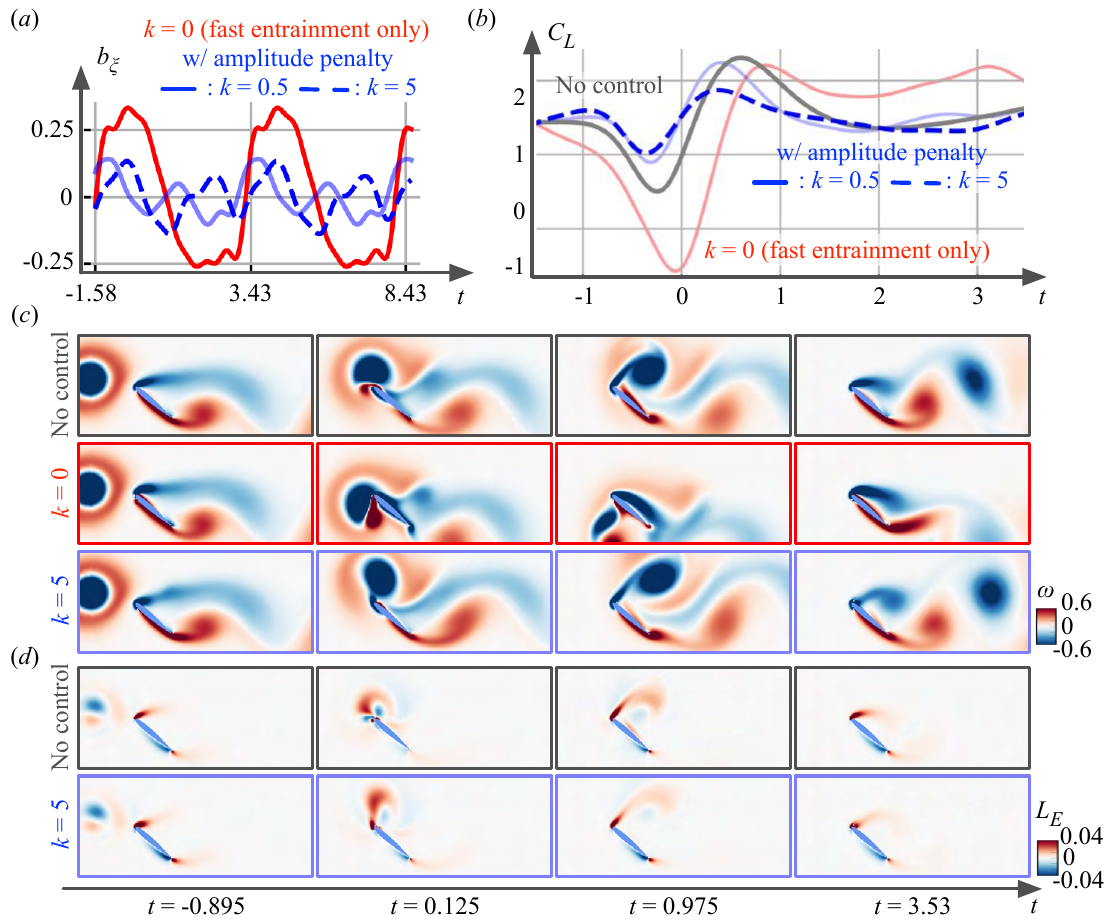}
    \vspace{-2mm}
    \caption{
    Phase-amplitude-based control of an extreme aerodynamic flow of $(\alpha, G,D,Y)=(40^\circ, -4, 0.5, 0.1)$.
    $(a)$ Optimal waveform ${b}_{\xi}$ with $k=0$, 0.5, and 5.
    $(b)$ Lift coefficient $C_L$ of the uncontrolled and controlled cases with $k=0$, 0.5, and 5.
    $(c)$ Vorticity fields and $(d)$ lift force elements of the uncontrolled case and the controlled cases with $k=0$ and 5.
    }
    % \vspace{-10mm}
    \label{fig13}
\end{figure}

Let us further examine the control effect with vorticity snapshots, as summarized in figure~\ref{fig11}$(c)$.
While the actuation at the leading edge already affects the vortical flows at $t=-0.895$, the effect on the vortex gust is clearly observed at $t=0.125$.
For $k=0$, the vortex core is shifted up due to the actuation, resulting in strong interaction with the leading-edge vortex at $t=0.975$.
This largely contributes to the amplification of lift response.
We note that a vorticity field at $t=3.53$ with $k=0$ presents more distinct rolled-up leading and trailing edge vortices that are observed at a high frequency~\cite{godavarthi2023optimal}.
This suggests that the fast synchronization-focused optimal waveform can quickly modify the flow states to be a target frequency.

In contrast, the amplitude-constrained optimal actuation with $k=5$ shifts the vortex core downward and the vortex gust moves to the pressure side of the airfoil.
Because of this modification of the vortex-core trajectory, the strong vortex gust merges with the trailing-edge vortex, as seen at $t=0.975$.
The wake behavior then eventually returns to the baseline natural vortex shedding.
In other words, the control strategy developed in a low-order space to suppress the amplitude modulation while quickly modifying the low-order dynamics works well to mitigate the gust impact in the high-dimensional physical space.
{Similar control performance has been confirmed when choosing $\omega_f>1$ and $k\approx 5$.}

To further analyze the aforementioned control effect, we perform the force element analysis~\cite{chang1992potential} which identifies responsible vortical structures for lift generation.
Let us define an auxiliary potential function $\phi_L$ with the boundary condition $-{\bm n}\cdot \nabla \phi_L = {\bm n}\cdot {\bm e}_y$ on the wing surface.
Here, ${\bm n}$ is the unit wall normal vector and ${\bm e}_y$ is the unit vector in the lift direction.
By taking the inner product of the Navier--Stokes equations with $\nabla \phi_L$ and performing an integral over the fluid domain, the lift force $F_L$ can be expressed as
\begin{align}
    F_L = \int_{\cal D} {\bm \omega}\times {\bm u}\cdot \nabla \phi_L dD + \dfrac{1}{Re}\int_{\partial D} {\bm \omega}\times {\bm n}\cdot (\nabla \phi_L + {\bm e}_y)dl,
\end{align}
where the first and second terms correspond to the surface integral and the line integral on the wing surface, respectively.
The first term is called the lift element $L_E ({\bm x},t)$, which has often been used to infer the source of lift generation in vortical flows~\cite{moriche2021characterization,zhang2022low,ribeiro2022wing,menon2022contribution}.

Lift element fields over extreme vortex-airfoil interaction are shown in figure~\ref{fig11}$(d)$.
For the uncontrolled case, the impingement of the vortex gust at {$t=0.125$} greatly contributes to the large transient lift force.
It is also observed that the interaction between the vortex gust and the separated leading edge vortex provides positive contribution to lift, which would be difficult to assess from vorticity fields only.

The mechanism of fluctuation suppression with the present control can be understood with the lift element analysis.
As shown, the downward shift of vortex core at {$t=0.125$} significantly reduces the positive contribution to the lift force.
In addition, at {$t=0.975$}, the positive vorticity structure generated due to the merging of the vortex gust and the trailing-edge vortex near the pressure side of the airfoil exhibits a negative effect on the lift force.
These suggest that the shift of the vortex gust with the present control can indeed reduce the lift force.

Let us also apply the present control to the case of $(\alpha, G,D,Y)=(40^\circ, -4, 0.5, 0.1)$ to examine the applicability for which a strong negative vortex impinges a wing.
In contrast to the case with positive vortex gusts, the lift force first drops and then increases.
The ratio between the target and natural frequencies is set to be $\omega_{\bm f}/\omega_{\alpha}=1.3$.
The optimal waveform ${\bm b}_{\bm \xi}$, the corresponding lift response, and vorticity fields are shown in figures~\ref{fig13}$(a)$--$(c)$.
For $k=0$, the lift fluctuation is rather amplified since the actuation is designed for fast synchronization only. 
By introducing the amplitude penalty, the lift fluctuation can be greatly suppressed, reporting $\eta = -0.410$.

Around $t=-0.8$, the lift force with the amplitude penalty is increased by the actuation.
This leads to the cancellation effect for the first drop of the lift force around $t= -0.3$, thereby contributing to the suppression of lift fluctuation.
This can also be evident from the shift up of vortex core occurring around the leading edge in a vorticity field at $t=0.125$.
Due to this vortex-core shift, the positive contribution to lift is enhanced at the leading edge, as shown in the lift force element field of figure~\ref{fig13}$(d)$.
Hence, the positive lift generation here contributes to canceling the reduction of lift force by a strong vortex gust.
% While the mechanism for controlling the flows would be different depending on vortex-gust parameters, 
These observations suggest the possibility of quick flow modification under extreme aerodynamic conditions with local actuation.

\section{Concluding remarks}
\label{sec:conc}

We presented a data-driven approach to mitigate the impact of vortex gusts for flows around an airfoil.
In particular, our consideration lies in the conditions of gust ratio $G>1$ that are challenging to sustain stable flights, referred to as extreme aerodynamics.
The present control strategy was developed in a low-dimensional manifold discovered by a nonlinear autoencoder.
Once a collection of extreme aerodynamic data is compressed into a three-dimensional latent space, we modeled the dynamics of the latent variables using a sparsity-promoting regression.
The identified dynamics as a form of ODE was used to perform phase-amplitude reduction, providing the phase and amplitude sensitivity functions.
These functions reveal the system sensitivity in terms of the phase shift and amplitude modulation against a given force input.
To quickly suppress the lift fluctuation of extreme vortex gust-airfoil interaction, the control actuation was derived through the amplitude-constrained optimal waveform analysis with the derived phase and amplitude sensitivity functions.
We found that the present control technique suppresses lift fluctuation due to a strong vortex impingement within a very short time for a wide variety of scenarios.
Furthermore, the successful impact mitigation with a localized forcing implies the possibility of gust control without necessitating drastic pitching motion of the wing.
{While additional investigations are needed, the proposed data-driven approach may be able to incorporate synthetic jet or plasma actuator-based active flow control strategies~\cite{greenblatt2022flow}.}

The present observations suggest the importance of physically-tractable data compression and preparation of appropriate coordinates to represent complex aerodynamic fluid flow data.
The present data compression approach enabled the use of SINDy to model the high-dimensional dynamics in a low-order manner.
{Furthermore, the compression reduces computational costs in deriving control techniques compared to conventional phase-based analyses performed in the original high-dimensional space, while the preparation for training data and the model development requires substantial effort.}
The present coherent low-order expression provides a connection between extreme aerodynamic flows and phase-amplitude reduction, enabling the analysis of seemingly complex fluid flows.
{While the direct application of the present method to higher Reynolds-number flows may be challenging, one may be able to consider topology-inspired nonlinear machine-learning-based compression that provides phase-amplitude-based coordinates even for non-periodic flows~\cite{smith2024cyclic}.}

% The reviewer is also correct for the second point about computational costs.
% The preparation for training data and the model training are indeed not cheap; however, designing control strategies with Floquet analysis on a low-order manifold enables significant reduction of computational costs.
% Furthermore, compared to the previous studies that perform phase-based control design in the original high-dimensional space (e.g., Kawamura et al., 2022, and Godavarthi et al., 2023), one challenge in this study is to find low-order coordinates that poses phase and amplitude representations for extreme vortex-airfoil interactions about the baseline dynamics, which can be achieved with the present lift-augmented autoencoder.
% Since the present phase-amplitude-based modeling can only be performed on such coordinates, we emphasize that preparing appropriate coordinates in a physically tractable manner is essential to perform phase-amplitude-based analysis for extreme aerodynamic flows in addition to computational cost reduction due to data compression.

There are some conceivable extensions of the present study.
One can consider the use of feedback formulation in designing the optimal control actuation in either the SINDy-based model space~\cite{brunton2015closed,nair2018networked} or waveform construction~\cite{takata2021fast}.
This could extend the control bounds of the present formulation for high gust ratio and large vortex gusts.
{While the present study considered library regression-based dynamical modeling whose explicit forms can often be examined~\cite{he2023pressure,FGT2024}, other candidates such as neural ODE~\cite{linot2023turbulence} and recurrent networks~\cite{SGASV2019} can be used for low-order dynamical modeling.}
{The dependence of the control performance on the shape and form given as the control input would also be of interest.} 
In addition, the present formulation can be combined with data-driven sparse reconstruction techniques as demonstrated with decoder-type neural-network-based efforts~\cite{erichson2020shallow,FukamiVoronoi,FFT2023_survey}.
{Towards real-time analysis that is essential for dealing with significant force deviation in a very short time, estimating low-dimensional extreme aerodynamic latent vectors from pressure sensors would be helpful~\cite{FT2024TSFP}.}
Since the proposed control can be performed in a three-dimensional latent space with minimal cost, the present idea of deriving optimal control actuation in a low-order space may prove useful for real-time stable flight operation of modern small-scale aircraft under extreme aerodynamic conditions.

\section*{Acknowledgements}
{KT and KF acknowledge the support from the US Air Force Office of Scientific Research (grant number: FA9550-21-1-0178) and the US Department of Defense Vannevar Bush Faculty Fellowship (grant number: N00014-22-1-2798).
KF thanks the support from the UCLA-Amazon Science Hub for Humanity and Artificial Intelligence.
HN acknowledges the support from JSPS KAKENHI (grant numbers: JP22K11919 and JP22H00516) and JST CREST (grant number: JP-MJCR1913). 
We thank Wataru Kurebayashi, Yuzuru Kato, Luke Smith, Alec Linot, and Vedasri Godavarthi for insightful discussions.
}

\section*{Declaration of interests}
{The authors report no conflict of interest.}

\section*{Appendix: Assessments of control performance for various extreme aerodynamic scenarios}

{
The present approach finds the optimal actuation based on phase and amplitude sensitivity functions derived from a SINDy-based low-order dynamical model.
However, the actuation pattern is designed in a feedforward manner.
Furthermore, phase-amplitude reduction involves some linear assumptions, implying that there may exist cases beyond the linear assumptions being valid.
Here, we examine the control performance of the present method for extreme vortex-airfoil interactions with various gust parameters to assess these matters.

The concept of phase-amplitude reduction could be leveraged for extreme aerodynamic cases that are mapped in the vicinity of the baseline limit cycle (similar to the case in figure~\ref{fig6}).
To quantify the deviation from the undisturbed dynamics, we consider the averaged distance $\Delta R_{\bm \xi}$ in a three-dimensional latent space,
\begin{align}
    \Delta R_{\bm \xi} = (\overline{R_{{\bm \xi},{\rm dist}}}- \overline{R_{{\bm \xi},{\rm base}}})/\overline{R_{{\bm \xi},{\rm base}}},
\end{align}
where 
\begin{align}
    R_{{\bm \xi},{\rm base}}^2 (t_i) = {\sum_j^3  (\xi_{j,{\rm base}}(t_i) - \overline{\xi_{j,{\rm base}}})^2},
    ~~~
    R_{{\bm \xi},{\rm dist}}^2 (t_i) = {\sum_j^3  (\xi_{j,{\rm dist}}(t_i) - \overline{\xi_{j,{\rm base}}})^2},
\end{align}
with the time-averaging operation $\overline{(\cdot)}$.
Hence, $R_{\bm \xi}$ measures the deviation from the baseline orbit in the latent space.

\begin{figure}
    \centering
    \includegraphics[width=\textwidth]{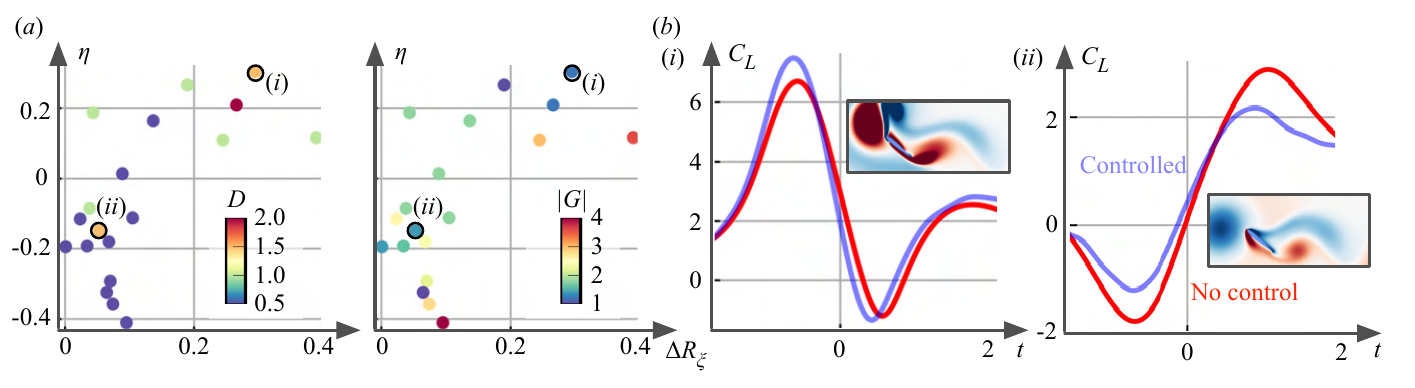}
    \vspace{-2mm}
    \caption{
    Assessments of the control bounds for extreme aerodynamic flows.
    $(a)$ The relationship between the control effect $\eta$ and the deviation of the latent vector from the undisturbed baseline state $\Delta R_{\bm \xi}$ colored by the vortex gust size $D$ and the absolute gust ratio $|G|$.
    $(b)$ Time series of lift coefficient $C_L$ for cases $(i)$ $(G,D,Y)=(3.6, 1, 0.1)$ and $(ii)$ $(-1.4, 1.5, 0)$ with uncontrolled snapshots are shown.
    }
    \label{fig16}
\end{figure}

The relationship between the control effect $\eta$ and the distance $\Delta R_{\bm \xi}$ is shown in figure~\ref{fig16}$(a)$.
The plots are colored by the size of the vortex disturbance $D$.
Here, we set the ratio between the natural and target frequencies to be $\omega_{\bm f}/\omega_\alpha=1.5$.
There is a clear trend --- the smaller $\Delta R_{\bm \xi}$, the better control performance (a negative $\eta$).
The present control is valid for cases that reside near the undisturbed baseline orbit in the latent space, as the effect of gusts may be considered as forcing through the sensitivity functions.
Since the dynamical behavior of the extreme aerodynamic trajectories on the manifold is generally affected by the vortex size more than the gust ratio~\cite{FT2023}, cases with a smaller gust size can be controlled relatively well.

There are also cases in which the lift fluctuation can be mitigated even with a large vortex gust.
We provide a lift curve for two extreme aerodynamic cases involving a large strong vortex gust with the parameters of $(i)$ $(G,D,Y)=(3.6, 1, 0.1)$ and $(ii)$ $(-1.4, 1.5, 0)$ in figure~\ref{fig16}$(b)$.
While the lift response of case $(i)$ does not show significant differences after the actuation due to large $G$ and $D$, case $(ii)$ with a larger gust size of $D=1.5$ achieves $15\%$ reduction of the lift fluctuation.
We should note that the gust ratio of $G=-1.4$ for case~$(ii)$ already belongs to the extreme condition.
These observations suggest that the present feedforward control strategy developed in a three-dimensional space may provide a step toward flying under extreme aerodynamic conditions.

}

%%%%%%%%%%%%%%%%%%%%%%%%%%%%%%%%
%%%%%%%%%%%%%%%%%%%%%%%%%%%%%%%%%
%%%%%%%%%%%%%%%%%%%%%%%%%%%%%%%%%

\bibliographystyle{unsrt}  
\bibliography{references}

\begin{thebibliography}{10}

\bibitem{cai2014survey}
G.~Cai, J.~Dias, and L.~Seneviratne.
\newblock A survey of small-scale unmanned aerial vehicles: {R}ecent advances
  and future development trends.
\newblock {\em Unmanned Syst.}, 2(02):175--199, 2014.

\bibitem{mishra2020drone}
B.~Mishra, D.~Garg, P.~Narang, and V.~Mishra.
\newblock Drone-surveillance for search and rescue in natural disaster.
\newblock {\em Comput. Commun.}, 156:1--10, 2020.

\bibitem{zhang2012application}
C.~Zhang and J.~M. Kovacs.
\newblock The application of small unmanned aerial systems for precision
  agriculture: a review.
\newblock {\em Precis. Agric.}, 13:693--712, 2012.

\bibitem{holton2015unmanned}
A.~E. Holton, S.~Lawson, and C.~Love.
\newblock Unmanned {A}erial {V}ehicles: {O}pportunities, barriers, and the
  future of ``drone journalism".
\newblock {\em Journal. Pract.}, 9(5):634--650, 2015.

\bibitem{jones2022physics}
A.~R. Jones, O.~Cetiner, and M.~J. Smith.
\newblock Physics and modeling of large flow disturbances: {D}iscrete gust
  encounters for modern air vehicles.
\newblock {\em Annu. Rev. Fluid Mech.}, 54:469--493, 2022.

\bibitem{mohamed2023gusts}
A.~Mohamed, M.~Marino, S.~Watkins, J.~Jaworski, and A.~Jones.
\newblock Gusts encountered by flying vehicles in proximity to buildings.
\newblock {\em Drones}, 7(1):22, 2023.

\bibitem{biler2021experimental}
H.~Biler, G.~Sedky, A.~R. Jones, M.~Saritas, and O.~Cetiner.
\newblock Experimental investigation of transverse and vortex gust encounters
  at low {R}eynolds numbers.
\newblock {\em AIAA J.}, 59(3):786--799, 2021.

\bibitem{stutz2023dimensional}
C.~M. Stutz, J.~T. Hrynuk, and D.~G. Bohl.
\newblock Dimensional analysis of a transverse gust encounter.
\newblock {\em Aerosp. Sci. Technol.}, 137:108285, 2023.

\bibitem{jones2021overview}
A.~R. Jones and O.~Cetiner.
\newblock Overview of unsteady aerodynamic response of rigid wings in gust
  encounters.
\newblock {\em AIAA J.}, 59(2):731--736, 2021.

\bibitem{FT2023}
K.~Fukami and K.~Taira.
\newblock Grasping extreme aerodynamics on a low-dimensional manifold.
\newblock {\em Nat. Commun.}, 14:6480, 2023.

\bibitem{qian2023lift}
Y.~Qian, Z.~Wang, and I.~Gursul.
\newblock Lift alleviation in travelling vortical gusts.
\newblock {\em Aeronaut. J.}, pages 1--22, 2023.

\bibitem{herrmann2022gust}
B.~Herrmann, S.~L. Brunton, J.~E. Pohl, and R.~Semaan.
\newblock Gust mitigation through closed-loop control. {II.} {F}eedforward and
  feedback control.
\newblock {\em Phys. Rev. Fluids}, 7(2):024706, 2022.

\bibitem{sedky2023experimental}
G.~Sedky, A.~Gementzopoulos, F.~D. Lagor, and A.~R. Jones.
\newblock Experimental mitigation of large-amplitude transverse gusts via
  closed-loop pitch control.
\newblock {\em Phys. Rev. Fluids}, 8(6):064701, 2023.

\bibitem{lumley1967structure}
J.~L. Lumley.
\newblock The structure of inhomogeneous turbulent flows.
\newblock {\em Atmospheric turbulence and radio wave propagation}, 1967.

\bibitem{berkooz1993proper}
G.~Berkooz, P.~Holmes, and J.~L. Lumley.
\newblock The proper orthogonal decomposition in the analysis of turbulent
  flows.
\newblock {\em Annu. Rev. Fluid Mech.}, 25(1):539--575, 1993.

\bibitem{Schmid2010}
P.~J. Schmid.
\newblock Dynamic mode decomposition of numerical and experimental data.
\newblock {\em J. Fluid Mech.}, 656:5--28, 2010.

\bibitem{BEF2019}
M.~P. Brenner, J.~D. Eldredge, and J.~B. Freund.
\newblock Perspective on machine learning for advancing fluid mechanics.
\newblock {\em Phys. Rev. Fluids}, 4:100501, 2019.

\bibitem{BHT2020}
S.~L. Brunton, M.~S. Hemanti, and K.~Taira.
\newblock Special issue on machine learning and data-driven methods in fluid
  dynamics.
\newblock {\em Theor. Comput. Fluid Dyn.}, 34(4):333--337, 2020.

\bibitem{BNK2020}
S.~L. Brunton, B.~R. Noack, and P.~Koumoutsakos.
\newblock Machine learning for fluid mechanics.
\newblock {\em Annu. Rev. Fluid Mech.}, 52:477--508, 2020.

\bibitem{nakao2021phase}
H.~Nakao.
\newblock Phase and amplitude description of complex oscillatory patterns in
  reaction-diffusion systems.
\newblock In {\em Physics of Biological Oscillators: New Insights into
  Non-Equilibrium and Non-Autonomous Systems}, pages 11--27. Springer, 2021.

\bibitem{shirasaka2017phase}
S.~Shirasaka, W.~Kurebayashi, and H.~Nakao.
\newblock Phase-amplitude reduction of transient dynamics far from attractors
  for limit-cycling systems.
\newblock {\em Chaos}, 27(2), 2017.

\bibitem{wedgwood2013phase}
K.~C.~A. Wedgwood, K.~K. Lin, R.~Thul, and S.~Coombes.
\newblock Phase-amplitude descriptions of neural oscillator models.
\newblock {\em J. Math. Neurosci.}, 3:1--22, 2013.

\bibitem{wilson2016isostable}
D.~Wilson and J.~Moehlis.
\newblock Isostable reduction of periodic orbits.
\newblock {\em Phys. Rev. E}, 94(5):052213, 2016.

\bibitem{YFTN2024}
K.~Yawata, K.~Fukami, K.~Taira, and H.~Nakao.
\newblock Phase autoencoder for limit-cycle oscillators.
\newblock {\em Chaos}, 34(6):063111, 2024.

\bibitem{mauroy2018global}
A.~Mauroy and I.~Mezi{\'c}.
\newblock Global computation of phase-amplitude reduction for limit-cycle
  dynamics.
\newblock {\em Chaos}, 28(7), 2018.

\bibitem{kotani2020nonlinear}
K.~Kotani, Y.~Ogawa, S.~Shirasaka, A.~Akao, Y.~Jimbo, and H.~Nakao.
\newblock Nonlinear phase-amplitude reduction of delay-induced oscillations.
\newblock {\em Phys. Rev. Res.}, 2(3):033106, 2020.

\bibitem{MZN2023}
P.~Mircheski, J.~Zhu, and H.~Nakao.
\newblock Phase-amplitude reduction and optimal phase locking of collectively
  oscillating networks.
\newblock {\em Chaos}, 33:103111, 2023.

\bibitem{nakao2016phase}
H.~Nakao.
\newblock Phase reduction approach to synchronisation of nonlinear oscillators.
\newblock {\em Contemp. Phys.}, 57(2):188--214, 2016.

\bibitem{kurebayashi2013phase}
W.~Kurebayashi, S.~Shirasaka, and H.~Nakao.
\newblock Phase reduction method for strongly perturbed limit cycle
  oscillators.
\newblock {\em Phys. Rev. Lett.}, 111(21):214101, 2013.

\bibitem{mauroy2013isostables}
A.~Mauroy, I.~Mezi{\'c}, and J.~Moehlis.
\newblock Isostables, isochrons, and {K}oopman spectrum for the action--angle
  representation of stable fixed point dynamics.
\newblock {\em Phys. D: Nonlinear Phenom.}, 261:19--30, 2013.

\bibitem{takeda2023two}
N.~Takeda, H.~Ito, and H.~Kitahata.
\newblock Two-dimensional hydrodynamic simulation for synchronization in
  coupled density oscillators.
\newblock {\em Phys. Rev. E}, 107(3):034201, 2023.

\bibitem{taira2018phase}
K.~Taira and H.~Nakao.
\newblock Phase-response analysis of synchronization for periodic flows.
\newblock {\em J. Fluid Mech.}, 846:R2, 2018.

\bibitem{iima2019jacobian}
M.~Iima.
\newblock Jacobian-free algorithm to calculate the phase sensitivity function
  in the phase reduction theory and its applications to {K}{\'a}rm{\'a}n's
  vortex street.
\newblock {\em Phys. Rev. E}, 99(6):062203, 2019.

\bibitem{khodkar2020phase}
M.~A. Khodkar and K.~Taira.
\newblock Phase-synchronization properties of laminar cylinder wake for
  periodic external forcings.
\newblock {\em J. Fluid Mech.}, 904:R1, 2020.

\bibitem{khodkar2021phase}
M.~A. Khodkar, J.~T. Klamo, and K.~Taira.
\newblock Phase-locking of laminar wake to periodic vibrations of a circular
  cylinder.
\newblock {\em Phys. Rev. Fluids}, 6(3):034401, 2021.

\bibitem{loe2021phase}
I.~A. Loe, H.~Nakao, Y.~Jimbo, and K.~Kotani.
\newblock Phase-reduction for synchronization of oscillating flow by
  perturbation on surrounding structure.
\newblock {\em J. Fluid Mech.}, 911:R2, 2021.

\bibitem{loe2023controlling}
I.~A. Loe, T.~Zheng, K.~Kotani, and Y.~Jimbo.
\newblock Controlling fluidic oscillator flow dynamics by elastic structure
  vibration.
\newblock {\em Sci. Rep.}, 13(1):8852, 2023.

\bibitem{iima2021phase}
M.~Iima.
\newblock Phase reduction technique on a target region.
\newblock {\em Phys. Rev. E}, 103(5):053303, 2021.

\bibitem{iima2023optimal}
M.~Iima.
\newblock Optimal external forces of the lock-in phenomena for flow past an
  inclined plate in uniform flow.
\newblock {\em Phys. Rev. E}, 109(4):045102, 2024.

\bibitem{nair2021phase}
A.~G. Nair, K.~Taira, B.~W. Brunton, and S.~L. Brunton.
\newblock Phase-based control of periodic flows.
\newblock {\em J. Fluid Mech.}, 927:A30, 2021.

\bibitem{asztalos2021modeling}
K.~J. Asztalos, S.~T.~M. Dawson, and D.~R. Williams.
\newblock Modeling the flow state sensitivity of actuation response on a
  stalled airfoil.
\newblock {\em AIAA J.}, 59(8):2901--2915, 2021.

\bibitem{kawamura2022adjoint}
Y.~Kawamura, V.~Godavarthi, and K.~Taira.
\newblock Adjoint-based phase reduction analysis of incompressible periodic
  flows.
\newblock {\em Phys. Rev. Fluids}, 7(10):104401, 2022.

\bibitem{godavarthi2023optimal}
V.~Godavarthi, Y.~Kawamura, and K.~Taira.
\newblock Optimal waveform for fast synchronization of airfoil wakes.
\newblock {\em J. Fluid Mech.}, 976:R1, 2023.

\bibitem{brunton2016discovering}
S.~L. Brunton, J.~L. Proctor, and J.~N. Kutz.
\newblock Discovering governing equations from data by sparse identification of
  nonlinear dynamical systems.
\newblock {\em Proc. Natl. Acad. Sci. U.S.A.}, 113(15):3932--3937, 2016.

\bibitem{takata2021fast}
S.~Takata, Y.~Kato, and H.~Nakao.
\newblock Fast optimal entrainment of limit-cycle oscillators by strong
  periodic inputs via phase-amplitude reduction and {F}loquet theory.
\newblock {\em Chaos}, 31(9), 2021.

\bibitem{cliff1}
F.~Ham and G.~Iaccarino.
\newblock Energy conservation in collocated discretization schemes on
  unstructured meshes.
\newblock In {\em Annual Research Briefs}, pages 3--14. Center for Turbulence
  Research, 2004.

\bibitem{cliff2}
F.~Ham, K.~Mattsson, and G.~Iaccarino.
\newblock Accurate and stable finite volume operators for unstructured flow
  solvers.
\newblock In {\em Annual Research Briefs}, pages 243--261. Center for
  Turbulence Research, 2006.

\bibitem{ZFAT2023}
Y.~Zhong, K.~Fukami, B.~An, and K.~Taira.
\newblock Sparse sensor reconstruction of vortex-impinged airfoil wake with
  machine learning.
\newblock {\em Theor. Comput. Fluid Dyn.}, 37:269--287, 2023.

\bibitem{kurtulus2015unsteady}
D.~F. Kurtulus.
\newblock On the unsteady behavior of the flow around {NACA} 0012 airfoil with
  steady external conditions at {$Re=1000$}.
\newblock {\em Int. J. Micro Air Veh.}, 7(3):301--326, 2015.

\bibitem{liu2012numerical}
Y.~Liu, K.~Li, J.~Zhang, H.~Wang, and L.~Liu.
\newblock Numerical bifurcation analysis of static stall of airfoil and dynamic
  stall under unsteady perturbation.
\newblock {\em Commun. Nonlinear Sci. Numer. Simul.}, 17(8):3427--3434, 2012.

\bibitem{di2018fluid}
G.~Di~Ilio, D.~Chiappini, S.~Ubertini, G.~Bella, and S.~Succi.
\newblock Fluid flow around {NACA} 0012 airfoil at low-{R}eynolds numbers with
  hybrid lattice {B}oltzmann method.
\newblock {\em Comput. Fluids}, 166:200--208, 2018.

\bibitem{taylor1918dissipation}
G.~I. Taylor.
\newblock On the dissipation of eddies.
\newblock {\em Meteorology, Oceanography and Turbulent Flow}, pages 96--101,
  1918.

\bibitem{HS2006}
G.~E. Hinton and R.~R. Salakhutdinov.
\newblock Reducing the dimensionality of data with neural networks.
\newblock {\em Science}, 313(5786):504--507, 2006.

\bibitem{fukami2021model}
K.~Fukami, K.~Hasegawa, T.~Nakamura, M.~Morimoto, and K.~Fukagata.
\newblock Model order reduction with neural networks: Application to laminar
  and turbulent flows.
\newblock {\em SN Comput. Sci.}, 2(467), 2021.

\bibitem{omata2019}
N.~Omata and S.~Shirayama.
\newblock A novel method of low-dimensional representation for temporal
  behavior of flow fields using deep autoencoder.
\newblock {\em AIP Adv.}, 9(1):015006, 2019.

\bibitem{xu2020multi}
J.~Xu and K.~Duraisamy.
\newblock Multi-level convolutional autoencoder networks for parametric
  prediction of spatio-temporal dynamics.
\newblock {\em Comput. Methods Appl. Mech. Eng.}, 372:113379, 2020.

\bibitem{racca2023predicting}
A.~Racca, N.~A.~K. Doan, and L.~Magri.
\newblock Predicting turbulent dynamics with the convolutional autoencoder echo
  state network.
\newblock {\em J. Fluid Mech.}, 975:A2, 2023.

\bibitem{smith2024cyclic}
L.~Smith, K.~Fukami, G.~Sedky, A.~Jones, and K.~Taira.
\newblock A cyclic perspective on transient gust encounters through the lens of
  persistent homology.
\newblock {\em J. Fluid Mech.}, 980:A18, 2024.

\bibitem{foias1988modelling}
C.~Foias, O.~Manley, and R.~Temam.
\newblock Modelling of the interaction of small and large eddies in two
  dimensional turbulent flows.
\newblock {\em ESAIM: Math. Model. Numer. Anal.}, 22(1):93--118, 1988.

\bibitem{temam1989inertial}
R.~Temam.
\newblock Do inertial manifolds apply to turbulence?
\newblock {\em Phys. D: Nonlinear Phenom.}, 37(1-3):146--152, 1989.

\bibitem{de2023data}
C.~E.~P. De~Jes{\'u}s and M.~D. Graham.
\newblock Data-driven low-dimensional dynamic model of {K}olmogorov flow.
\newblock {\em Phys. Rev. Fluids}, 8(4):044402, 2023.

\bibitem{FFT2019a}
K.~Fukami, K.~Fukagata, and K.~Taira.
\newblock Super-resolution reconstruction of turbulent flows with machine
  learning.
\newblock {\em J. Fluid Mech.}, 870:106--120, 2019.

\bibitem{wang2004image}
Z.~Wang, A.~C. Bovik, H.~R. Sheikh, and E.~P. Simoncelli.
\newblock Image quality assessment: from error visibility to structural
  similarity.
\newblock {\em IEEE Trans. Image Process.}, 13(4):600--612, 2004.

\bibitem{anatharaman2023image}
V.~Anantharaman, J.~Feldkamp, K.~Fukami, and K.~Taira.
\newblock Image and video compression of fluid flow data.
\newblock {\em Theor. Comput. Fluid Dyn.}, 37(1):61--82, 2023.

\bibitem{anderson1991fundamentals}
J.~D. Anderson.
\newblock {\em Fundamentals of {A}erodynamics, {McGraw-Hill}}.
\newblock New York, 1991.

\bibitem{sedky2020lift}
G.~Sedky, A.~R. Jones, and F.~D. Lagor.
\newblock Lift regulation during transverse gust encounters using a modified
  {G}oman--{K}hrabrov model.
\newblock {\em AIAA J.}, 58(9):3788--3798, 2020.

\bibitem{he2020stall}
G.~He, J.~Deparday, L.~Siegel, A.~Henning, and K.~Mulleners.
\newblock Stall delay and leading-edge suction for a pitching airfoil with
  trailing-edge flap.
\newblock {\em AIAA J.}, 58(12):5146--5155, 2020.

\bibitem{BPK2016b}
S.~L. Brunton, J.~L. Proctor, and J.~N. Kutz.
\newblock Sparse identification of nonlinear dynamics with control.
\newblock {\em IFAC NOLCOS}, 49(18):710--715, 2016.

\bibitem{KKB2018}
E.~Kaiser, J.~N. Kutz, and S.~L. Brunton.
\newblock Sparse identification of nonlinear dynamics for model predictive
  control in the low-data limit.
\newblock {\em Proc. Roy. Soc. A}, 474(2219):20180335, 2018.

\bibitem{LKLLBK2019}
S.~Li, E.~Kaiser, S.~Laima, H.~Li, S.~L. Brunton, and J.~N. Kutz.
\newblock Discovering time-varying aerodynamics of a prototype bridge by sparse
  identification of nonlinear dynamical systems.
\newblock {\em Phys. Rev. E}, 100(2):022220, 2019.

\bibitem{fukami2020sparse}
K.~Fukami, T.~Murata, K.~Zhang, and K.~Fukagata.
\newblock Sparse identification of nonlinear dynamics with low-dimensionalized
  flow representations.
\newblock {\em J. Fluid Mech.}, 926:A10, 2021.

\bibitem{Zou2006}
H.~Zou.
\newblock The adaptive lasso and its oracle properties.
\newblock {\em J. Am. Stat. Assoc.}, 101(476):1418--1429, 2006.

\bibitem{kuramoto1984chemical}
Y.~Kuramoto.
\newblock {\em Chemical oscillations, waves, and turbulence}.
\newblock Springer, 1984.

\bibitem{ermentrout1996type}
B.~Ermentrout.
\newblock Type {I} membranes, phase resetting curves, and synchrony.
\newblock {\em Neural Comput.}, 8(5):979--1001, 1996.

\bibitem{brown2004phase}
E.~Brown, J.~Moehlis, and P.~Holmes.
\newblock On the phase reduction and response dynamics of neural oscillator
  populations.
\newblock {\em Neural Comput.}, 16(4):673--715, 2004.

\bibitem{kuramoto2019concept}
Y.~Kuramoto and H.~Nakao.
\newblock On the concept of dynamical reduction: the case of coupled
  oscillators.
\newblock {\em Philos. Trans. R. Soc. A}, 377(2160):20190041, 2019.

\bibitem{harada2010optimal}
T.~Harada, H.~Tanaka, M.~J. Hankins, and I.~Z. Kiss.
\newblock Optimal waveform for the entrainment of a weakly forced oscillator.
\newblock {\em Phys. Rev. Lett.}, 105(8):088301, 2010.

\bibitem{zlotnik2016phase}
A.~Zlotnik, R.~Nagao, I.~Z. Kiss, and J.-S. Li.
\newblock Phase-selective entrainment of nonlinear oscillator ensembles.
\newblock {\em Nat. Commun.}, 7(1):10788, 2016.

\bibitem{zhang2022wake}
K.~Zhang and M.~N. Haque.
\newblock Wake interactions between two side-by-side circular cylinders with
  different sizes.
\newblock {\em Phys. Rev. Fluids}, 7(6):064703, 2022.

\bibitem{hoppensteadt1997weakly}
F.~C. Hoppensteadt and E.~M. Izhikevich.
\newblock {\em Weakly connected neural networks}.
\newblock Springer Science \& Business Media, 1997.

\bibitem{shim2007synchronized}
S.-B. Shim, M.~Imboden, and P.~Mohanty.
\newblock Synchronized oscillation in coupled nanomechanical oscillators.
\newblock {\em Science}, 316(5821):95--99, 2007.

\bibitem{zlotnik2013optimal}
A.~Zlotnik, Y.~Chen, I.~Z. Kiss, H.~Tanaka, and J.-S. Li.
\newblock Optimal waveform for fast entrainment of weakly forced nonlinear
  oscillators.
\newblock {\em Phys. Rev. Lett.}, 111(2):024102, 2013.

\bibitem{l-curve}
P.~C. Hansen and D.~P. O’Leary.
\newblock The use of the {L}-curve in the regularization of discrete ill-posed
  problems.
\newblock {\em SIAM J. Sci. Comput.}, 14(6):1487–1503, 1993.

\bibitem{chang1992potential}
C.-C. Chang.
\newblock Potential flow and forces for incompressible viscous flow.
\newblock {\em Proc. Roy. Soc. A}, 437(1901):517--525, 1992.

\bibitem{moriche2021characterization}
M.~Moriche, G.~Sedky, A.~R. Jones, O.~Flores, and M.~Garc{\'\i}a-Villalba.
\newblock Characterization of aerodynamic forces on wings in plunge maneuvers.
\newblock {\em AIAA J.}, 59(2):751--762, 2021.

\bibitem{zhang2022low}
K.~Zhang, B.~Shah, and O.~Bilgen.
\newblock Low-{R}eynolds-number aerodynamic characteristics of airfoils with
  piezocomposite trailing surfaces.
\newblock {\em AIAA J.}, 60(4):2701--2706, 2022.

\bibitem{ribeiro2022wing}
J.~H.~M. Ribeiro, C.-A. Yeh, K.~Zhang, and K.~Taira.
\newblock Wing sweep effects on laminar separated flows.
\newblock {\em J. Fluid Mech.}, 950:A23, 2022.

\bibitem{menon2022contribution}
K.~Menon, S.~Kumar, and R.~Mittal.
\newblock Contribution of spanwise and cross-span vortices to the lift
  generation of low-aspect-ratio wings: {I}nsights from force partitioning.
\newblock {\em Phys. Rev. Fluids}, 7(11):114102, 2022.

\bibitem{greenblatt2022flow}
D.~Greenblatt and D.~R. Williams.
\newblock Flow control for unmanned air vehicles.
\newblock {\em Annu. Rev. Fluid Mech.}, 54:383--412, 2022.

\bibitem{brunton2015closed}
S.~L. Brunton and B.~R. Noack.
\newblock Closed-loop turbulence control: {P}rogress and challenges.
\newblock {\em Appl. Mech. Rev.}, 67(5):050801, 2015.

\bibitem{nair2018networked}
A.~G. Nair, S.~L. Brunton, and K.~Taira.
\newblock Networked-oscillator-based modeling and control of unsteady wake
  flows.
\newblock {\em Phys. Rev. E}, 97(6):063107, 2018.

\bibitem{he2023pressure}
X.~He and D.~R. Williams.
\newblock Pressure feedback control of aerodynamic loads on a delta wing in
  transverse gusts.
\newblock {\em AIAA J.}, 61(4):1659--1674, 2023.

\bibitem{FGT2024}
K.~Fukami, S.~Goto, and K.~Taira.
\newblock Data-driven nonlinear turbulent flow scaling with {B}uckingham {P}i
  variables.
\newblock {\em J. Fluid Mech.}, 984:R4, 2024.

\bibitem{linot2023turbulence}
A.~J. Linot, K.~Zeng, and M.~D. Graham.
\newblock Turbulence control in plane couette flow using low-dimensional neural
  {ODE}-based models and deep reinforcement learning.
\newblock {\em Int. J. Heat Fluid Flow}, 101:109139, 2023.

\bibitem{SGASV2019}
P.~A. Srinivasan, L.~Guastoni, H.~Azizpour, P.~Schlatter, and R~Vinuesa.
\newblock Predictions of turbulent shear flows using deep neural networks.
\newblock {\em Phys. Rev. Fluids}, 4:054603, 2019.

\bibitem{erichson2020shallow}
N.~B. Erichson, L.~Mathelin, Z.~Yao, S.~L. Brunton, M.~W. Mahoney, and J.~N.
  Kutz.
\newblock Shallow neural networks for fluid flow reconstruction with limited
  sensors.
\newblock {\em Proc. Roy. Soc. A}, 476(2238):20200097, 2020.

\bibitem{FukamiVoronoi}
K.~Fukami, R.~Maulik, N.~Ramachandra, K.~Fukagata, and K.~Taira.
\newblock Global field reconstruction from sparse sensors with {Voronoi}
  tessellation-assisted deep learning.
\newblock {\em Nat. Mach. Intell.}, 3:945--951, 2021.

\bibitem{FFT2023_survey}
K.~Fukami, K.~Fukagata, and K.~Taira.
\newblock Super-resolution analysis via machine learning: a survey for fluid
  flows.
\newblock {\em Theor. Comput. Fluid Dyn.}, 37:421--444, 2023.

\bibitem{FT2024TSFP}
K.~Fukami and K.~Taira.
\newblock Extreme aerodynamics of vortex impingement: {M}achine-learning-based
  compression and situational awareness.
\newblock {\em 13rd International Symposium on Turbulence and Shear Flow
  Phenomena (TSFP13), Montréal, Canada}, (114), 2024.

\end{thebibliography}
%%% Remove comment to use the external .bib file (using bibtex).

\end{document}